\documentclass[aps,preprintnumbers,nofootinbib,superscriptaddress,showpacs,letterpaper,twocolumn]{revtex4-1}
\pdfoutput=1
\usepackage{amsmath,amssymb,amsbsy}
\usepackage[utf8]{inputenc}
\usepackage{graphicx,booktabs,}
\usepackage{ulem,color,topcapt}
\usepackage{comment}
\usepackage[colorlinks=true, linkcolor=blue, filecolor=blue, urlcolor=blue, citecolor=blue, pdftex, plainpages=false]{hyperref}

\definecolor{orange}{rgb}{1.0, 0.5, 0}

\begin{document}
\title{Dislocations under gradient flow and their effect on the renormalized coupling}
\author{Anna Hasenfratz}
\email{anna.hasenfratz@colorado.edu}
\affiliation{Department of Physics, University of Colorado, Boulder, Colorado 80309}
\author{Oliver Witzel}
\email{oliver.witzel@colorado.edu}
\affiliation{Department of Physics, University of Colorado, Boulder, Colorado 80309}

\date{\today}

\begin{abstract}
  Non-zero topological charge is prohibited in the chiral limit of gauge-fermion systems  because any instanton would create a zero mode of the Dirac operator.  On the lattice, however, the geometric $Q_\text{geom}=\langle F{\tilde F}\rangle /32\pi^2$  definition of the topological charge does not necessarily vanish even when the gauge fields are smoothed for example with gradient flow. Small vacuum fluctuations (dislocations) not seen by the fermions  may be promoted  to instanton-like objects by the gradient flow.
 We  demonstrate that these  artifacts of the flow cause the gradient flow renormalized gauge coupling to  increase and run faster.  In step-scaling studies  such artifacts contribute a term  which increases with  volume. The usual $a/L\to 0$ continuum limit extrapolations can hence lead to incorrect results.   
In this paper we investigate these topological lattice artifacts in the SU(3) 10-flavor system with domain wall fermions and the 8-flavor system with staggered fermions. Both  systems exhibit  nonzero topological charge  at the strong coupling, especially when using Symanzik gradient flow. We demonstrate   how this artifact  impacts the determination of the renormalized gauge coupling and the step-scaling $\beta$ function.
\end{abstract}
\maketitle

\section{Introduction}

The net instanton charge $Q=n_+-n_-$ is a topologically protected quantity in continuum gauge-fermion systems.  On the lattice, however, $Q$ is not protected and different definitions of the topological charge, like the number of zero modes of the Dirac operator
\begin{equation}
Q_\text{ferm} = \frac{1}{2} \text{tr} (\gamma_5 D),
\label{eq:Q_ferm}
\end{equation}
 or the geometric definition 
 \begin{equation}
 Q_\text{geom} =\frac{1}{32\pi^2} \int dx \; \text{tr} ( F_{\mu\nu}(x){\tilde F_{\mu\nu}}(x) )
 \label{eq:Q_geom}
 \end{equation}
may not agree. The latter definition is particularly troublesome as the ultraviolet (UV) fluctuations of the gauge field can dominate $Q_\text{geom}$. Smoothing the gauge field with smearing or gradient flow (GF)  reduces the problem, but the fate of small instanton-like objects, dislocations,  depends on the details. These  may grow to  topological modes $|Q|\approx 1$  with some but not with  other definitions of $Q_\text{geom}$. There is no unique definition of the topological charge on the lattice. Different definitions are expected to agree only in the continuum limit \cite{Schafer:1996wv,DeGrand:2000gq,DeGrand:2002vu,Bernard:2003gq,DelDebbio:2004ns,Durr:2006ky,Bruno:2014ova,Aoki:2017paw}.

Dynamical configurations with net topological charge $Q_\text{ferm} $ have $|Q_\text{ferm} |$ zero modes in the spectrum of the massless Dirac operator  assuming the lattice fermions are chirally symmetric \cite{Leutwyler:1992yt,Hasenfratz:1998ri}.  As a consequence, non-zero $Q_\text{ferm} $ configurations are excluded  in the chiral limit.  This is one of the rare instances  where theoretical arguments rigorously constrain the value of $Q$ even at finite cutoff.\footnote{Residual chiral symmetry breaking or effects due to the finite volume could potentially allow configurations with nonzero topological charge. This effect is however negligible.} The fermions restrict $Q_\text{ferm}=0$ and any $Q_\text{geom}\ne 0$ signals a lattice artifact of the smoothing algorithm or the operator used in the definition of $Q_\text{geom}$.  Even though it is a lattice artifact, $Q_\text{geom}\ne 0$ can have significant effects on the renormalized gradient flow gauge coupling and the finite volume step scaling function $\beta_{c,s}(g^2_{c})$. 

The GF gauge coupling at energy scale $\mu = 1/\sqrt{8t}$ is $g^2_{GF}(t;L,\beta) = \mathcal{N} t^2 \langle E \rangle$ where $\langle E \rangle$ is the energy density,  $\beta=6/g^2_0$ is the bare coupling,  $L$ refers to the linear size of the system, and the normalization factor $ \mathcal{N}$ is chosen to match $g^2_{\overline {MS}}$ at one-loop \cite{Luscher:2009eq,Luscher:2010iy,Narayanan:2006rf}.
Lattice studies show that at large flow time $g^2_{GF}(t)$  exhibits only mild, approximately linear or weaker, dependence on $t$. Therefore the energy density $\langle E \rangle$ decreases  $\propto 1/t$ or faster. While GF removes vacuum fluctuations and instanton pairs, some instantons can survive the flow and become (quasi-)stable.  At large flow time $Q_\text{geom}$ approaches integer values and $\langle Q_\text{geom}^2\rangle$ is frequently used to  define  the lattice topological susceptibility \cite{Bruno:2014ova,Aoki:2017paw,Alexandrou:2015yba,Alexandrou:2017hqw}. To simplify the notation we will from now on refer to $Q_\text{geom}$ simply by using $Q$.

The action of a single instanton is $S_I=8\pi^2$ in the continuum. On the lattice this value depends on the instanton size and the lattice action, but at large flow time smooth instantons  increase the energy of the configuration by  $\approx S_I$~\cite{DeGrand:2002vu}. The net number of instantons $Q=n_+-n_-$  is expected to scale with the square root of the number of lattice sites  $V/a^4$,  even if they arise from vacuum fluctuations as  artifacts of the GF.  The instanton  contribution  to the energy density is therefore $\propto 1/\sqrt{a^4V}$.  If  instanton-antiinstanton pairs are present, this contribution is even larger.

In step scaling studies the GF flow time is connected to the lattice size as $t=(c L)^2/8$, where the constant $c$ defines the renormalization scheme and $g_c^2$ refers to the gradient flow renormalized coupling at the corresponding flow time  $g_{GF}^2(t;L,\beta)$. The discrete lattice $\beta$ function  of scale change $s$ is defined as~\cite{Fodor:2012td}
\begin{equation}
\beta_{c,s}(g^2_c;L,\beta) = \frac{g^2_c(sL; \beta)- g^2_c(L; \beta)}{\log\;s^2} \,.
\end{equation}
 In volumes $V=L^4$  the contribution of the instantons to the discrete $\beta$ function  is $\propto t^2 /\sqrt{a^4V} \propto L^2/a^2$,
\begin{equation}
\beta_{c,s}(g^2_c;L,\beta) = \beta_{c,s}(g^2_c)_{Q=0} + \mathcal{C}(\beta,c) L^2/a^2, 
\label{eq:gGF_Q}
\end{equation}
where  $ \beta_{c,s}(g^2_c)_{Q=0}$ is the step scaling function in the $Q=0$ sector and $\mathcal{C}$ depends on the bare coupling $\beta$ and the renormalization scheme $c$ but is independent of the lattice size $L$.  When the simulations are performed with chirally symmetric fermions in the chiral limit, the term $\mathcal{C}(\beta,c) L^2/a^2$ is a lattice artifact, the consequence of the GF promoting vacuum fluctuations  to  topological objects.

Even on $Q=0$ configurations $\beta_{c,s}(g^2_c)_{Q=0} $ has cutoff effects. These are typically removed by an $a^2/L^2\to 0$ extrapolation at fixed renormalized coupling $g^2_c$ \cite{Fodor:2012td,Fodor:2014cpa,Hasenfratz:2014rna,DallaBrida:2016kgh,Ramos:2015baa,Hasenfratz:2016dou,Hasenfratz:2017qyr,Hasenfratz:2019dpr}. If the data does not follow  $a^2/L^2$ dependence, higher order $(a/L)^4$ terms  can be included \cite{Fodor:2017gtj,Fodor:2019ypi}. However, in the strong coupling limit with a non-negligible instanton density, Eq.~(\ref{eq:gGF_Q}) implies that  the correct continuum extrapolation should include an $(L/a)^2$ term instead or at least in addition to $(a/L)^4$.  Practically such an  $L\to\infty$ extrapolation  is not viable. This is a reflection of the 
non-perturbative nature of instantons and shows that their effect cannot be removed by perturbatively motivated extrapolations. 
The effect of instanton-like objects   in the continuum prediction could be substantial,  especially in slowly running systems near or within the conformal window where the coupling $\beta=6/g^2_0$ barely changes as the continuum limit is taken on available lattice volumes. A clean way to  avoid this issue is to choose a flow where instantons are not generated even on coarse lattices.

In this paper we re-analyze simulations performed  at strong coupling where the gauge fields are rough and dislocations  frequent.  Such simulations are necessary to explore the step scaling function of (near) conformal systems.  This phenomena might also affect scale setting \cite{Luscher:2010iy,Borsanyi:2012zs,Sommer:2014mea} in strongly coupled beyond the Standard Model systems (see e.g.~\cite{Witzel:2019jbe,Brower:2019oor,Fodor:2012ty,Fodor:2015zna,Fodor:2016wal,Appelquist:2014zsa,Appelquist:2018yqe,Aoki:2016wnc,Aoki:2013xza,Brower:2014ita,Brower:2015owo,Hasenfratz:2016gut,Witzel:2018gxm,Witzel:2019oej}) or even quantum chromodynamics (QCD) simulations at coarse lattice spacings necessary to achieve a large physical box needed e.g.~to study multi-particle interactions \cite{Detmold:2019ghl}.

We consider two different systems to illustrate the  issue. In both cases we study two different gradient flow kernels, Wilson and Symanzik flow. We start with our recent 10-flavor SU(3) domain wall simulations where we first observed  the effect of non-zero topological charge \cite{Hasenfratz:2017qyr}. An accompanying paper discusses the step-scaling function of this most likely conformal system  and provides further details \cite{Nf10stepScaling}. Next we analyze configurations generated for an older study of the SU(3) 8-flavor system with staggered fermions \cite{Hasenfratz:2014rna}.  We chose these two systems because both simulations have been pushed toward very strong coupling where the  contamination from topological modes  can be significant. Our results demonstrate these lattice artifacts are more severe for Symanzik than for Wilson flow. In Section \ref{Sec.Tflow} we demonstrate how a small modification of the flow  kernel results in a gradient flow that is better at smoothing out local dislocations resulting in fewer configurations with nonzero topological charge.  The lattice discretization errors of such a modified gradient flow will need to be explored in the future. Finally we briefly summarize our findings.

\section{\texorpdfstring{SU(3) with $N_f=10$ flavors}{SU(3) with Nf=10 flavors}}
\label{Sec.Nf10}
\subsection{Details of the simulations}
In this part of our study we utilize existing gauge field configurations generated with ten degenerate and massless flavors of three times stout-smeared \cite{Morningstar:2003gk} M\"obius domain wall (DW) fermions \cite{Shamir:1993zy,Furman:1994ky,Brower:2012vk} with Symanzik gauge action \cite{Luscher:1985zq,Luscher:1984xn}. The configurations are generated using \texttt{Grid}\footnote{\url{https://github.com/paboyle/Grid}} \cite{Boyle:2015tjk} and we choose symmetric volumes with $V=L^4$ where the gauge fields have periodic, the fermions antiperiodic boundary conditions in all four space-time directions. The bare input quark mass is zero and for the domain wall fermions we choose the domain wall height $M_5=1$ and the extent of the fifth dimension $L_s=16$. Configurations are generated using the hybrid Monte Carlo update algorithm \cite{Duane:1987de} choosing trajectories of length two molecular time units (MDTU) and we use configurations saved every 10 MDTU. Our statistical data analysis is performed using the $\Gamma$-method \cite{Wolff:2003sm} which estimates and accounts for integrated autocorrelation times. For the $L/a=32$ ensembles at strong coupling considered here autocorrelations range from three to five measurements.

Due to the finite extent of the fifth dimension, DW fermions exhibit a small, residual chiral symmetry breaking which conventionally is parametrized by an additive mass term $am_\text{res}$. We determine $am_{res}$ numerically using the ratio of midpoint-pseudoscalar and pseudoscalar-pseudoscalar correlator. At strong coupling $am_\text{res}$ depends on the bare coupling $\beta$ and increases from $am_\text{res} = 2\times 10^{-5}$ at $\beta=4.15$ to $6\times 10^{-4}$ at $\beta=4.02$. To demonstrate that $am_\text{res}$ is sufficiently small and \emph{not}\/ the origin of nonzero topological charges, we compare results for $\beta=4.05$ from ensembles with $L_s=16$ and $L_s=32$ below.

\subsection{Effects of nonzero topological charge}
\begin{figure*}[p]
  \includegraphics[width=0.95\columnwidth]{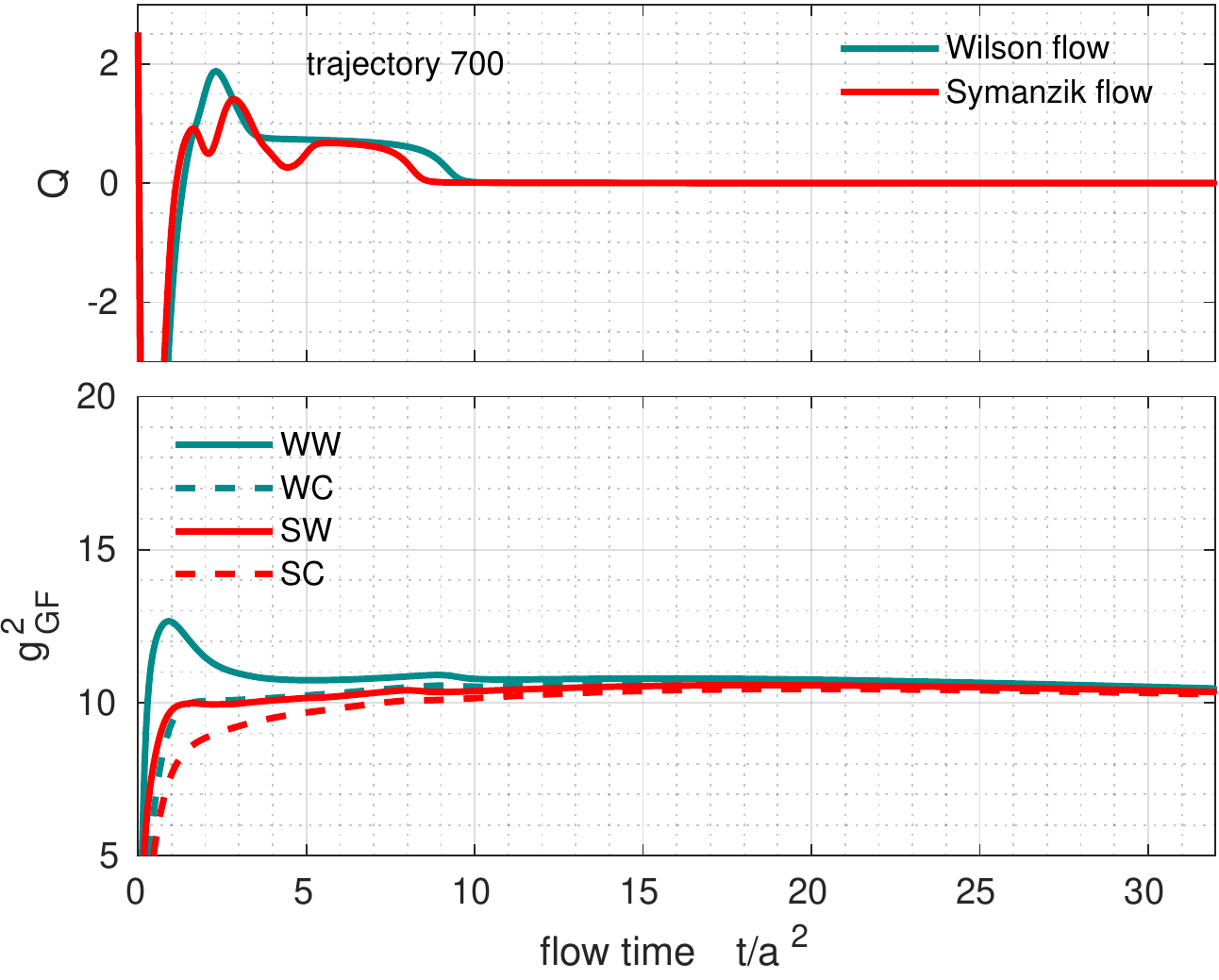}\hfill
  \includegraphics[width=0.95\columnwidth]{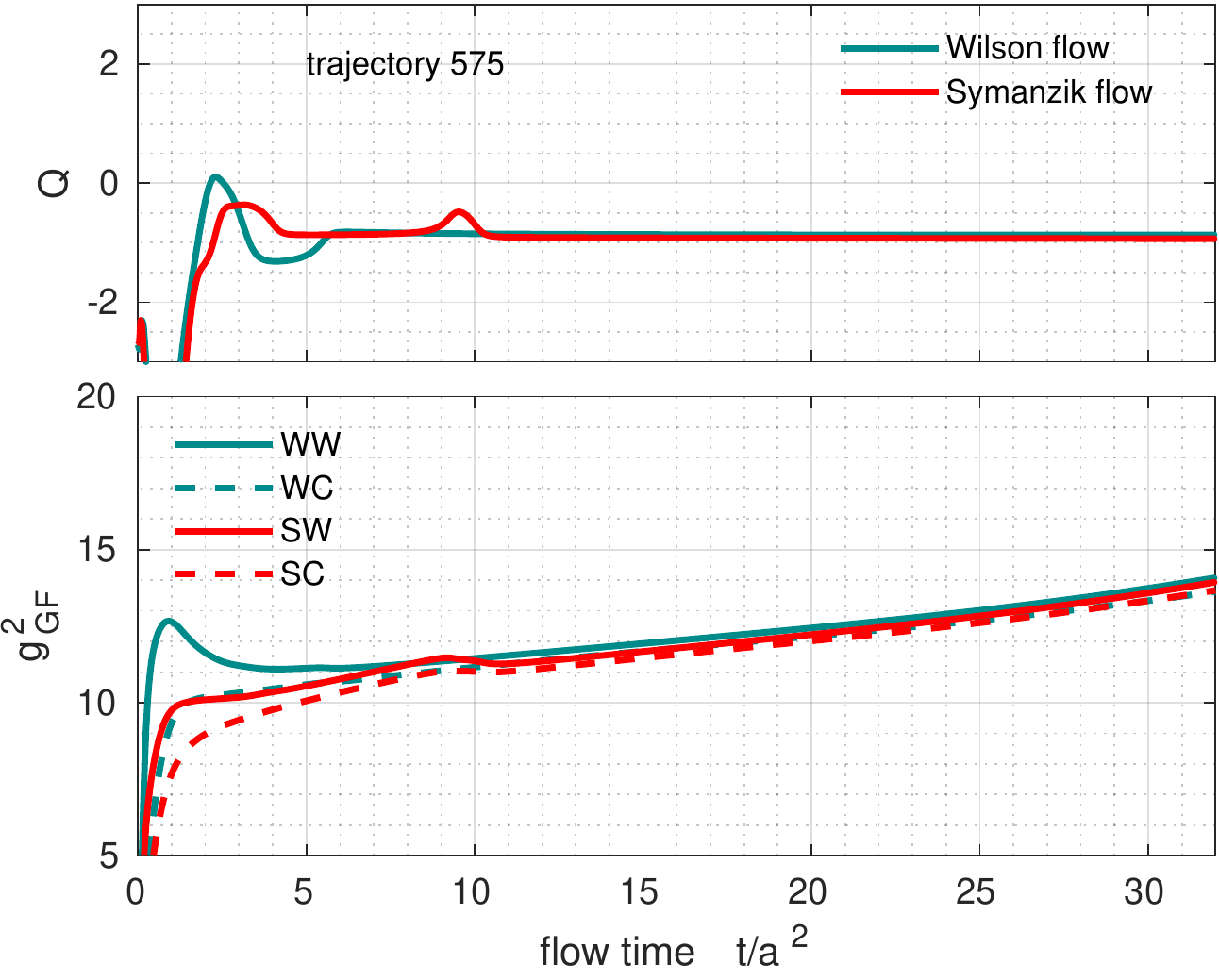}\\ \vspace{5mm}
  \includegraphics[width=0.95\columnwidth]{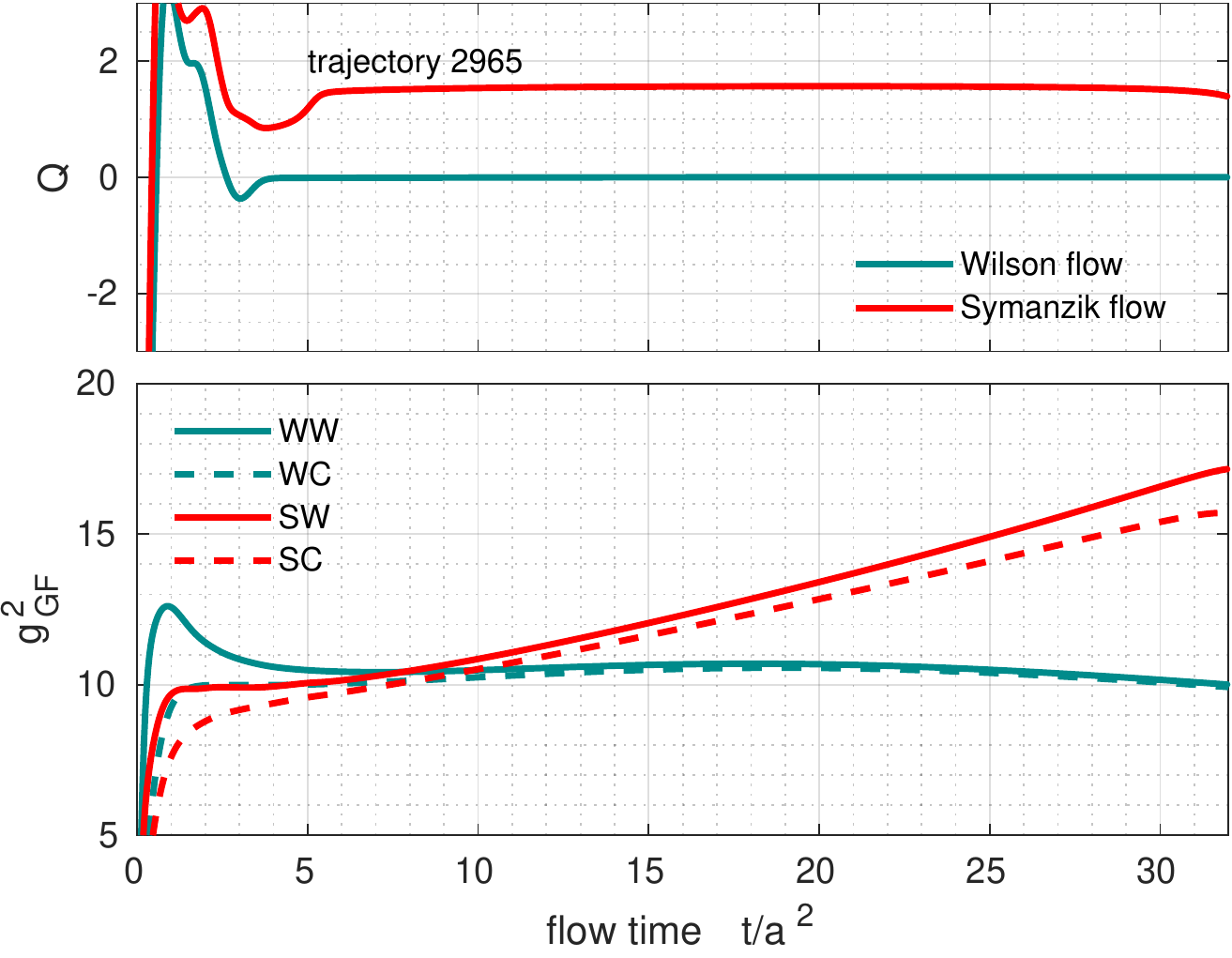} \hfill
  \includegraphics[width=0.95\columnwidth]{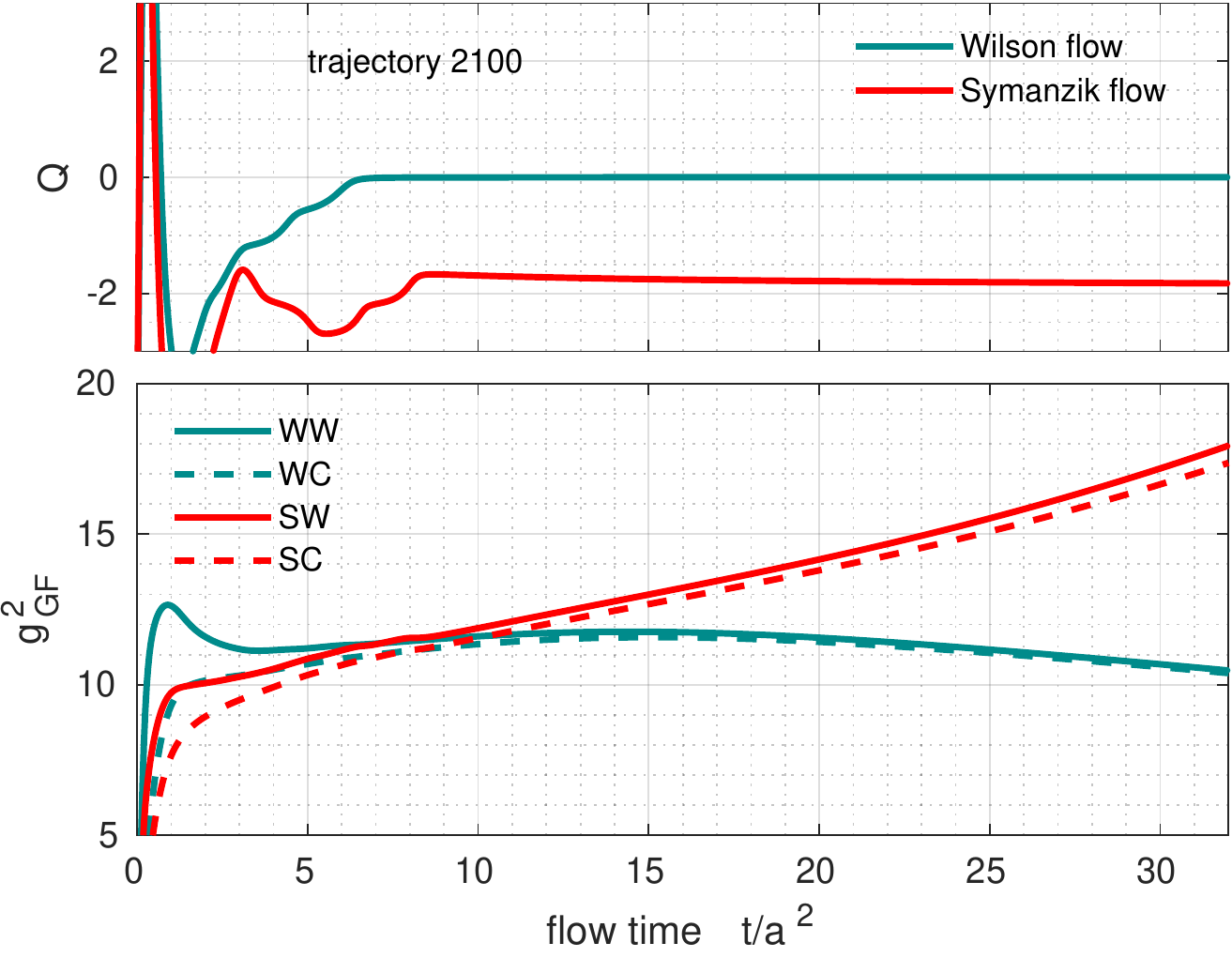} \\ \vspace{5mm}
  \includegraphics[width=0.95\columnwidth]{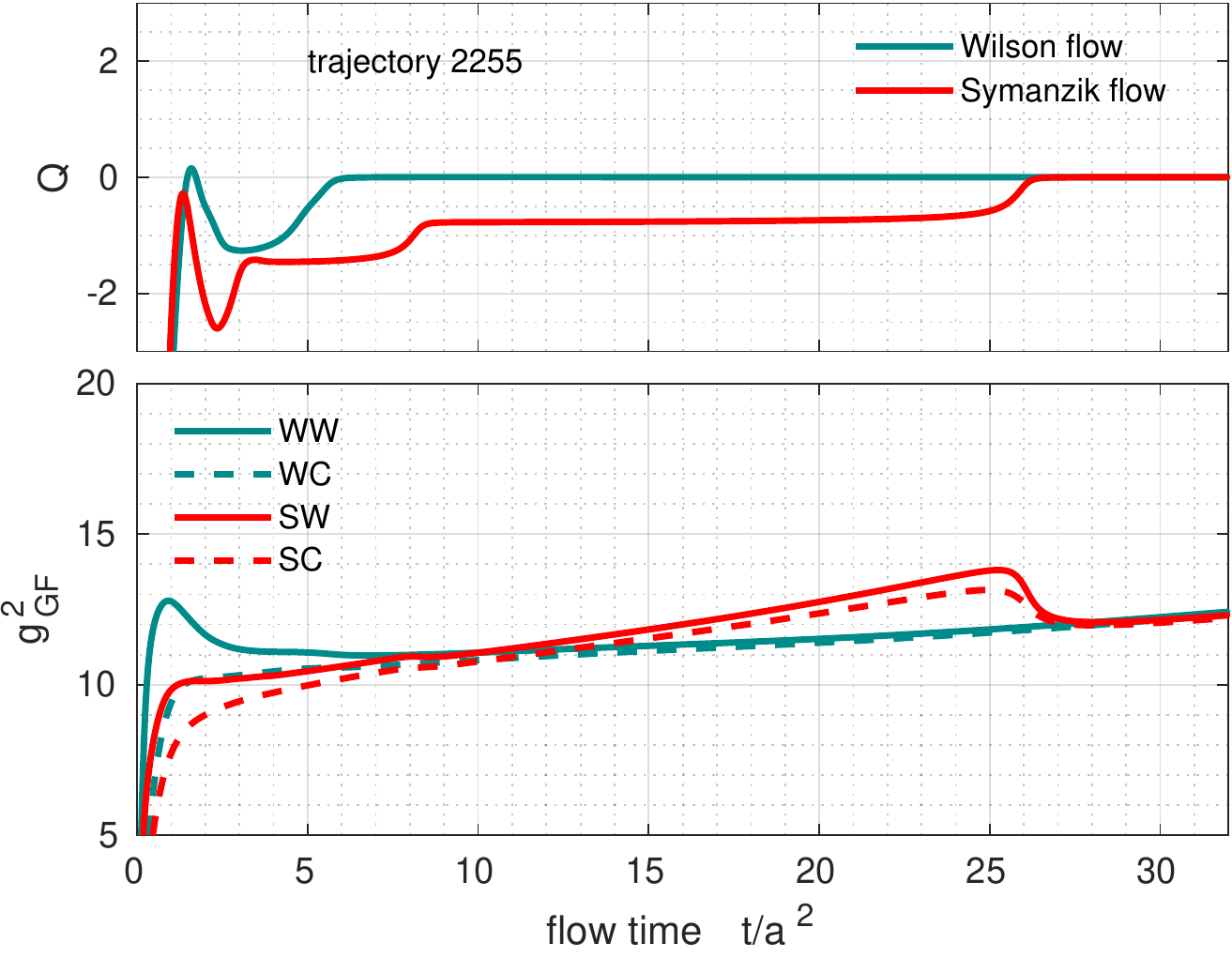}\hfill  
  \includegraphics[width=0.95\columnwidth]{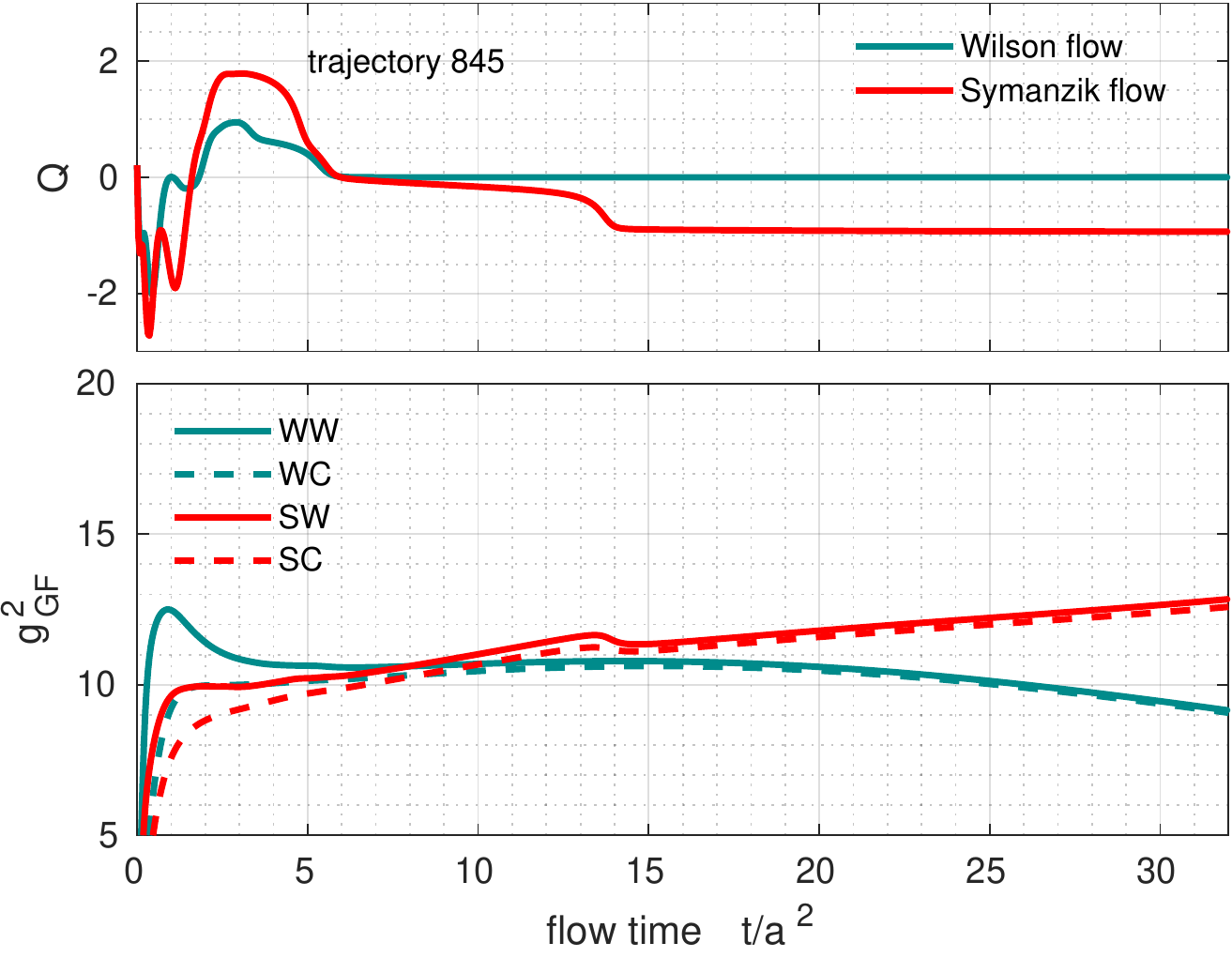}
\caption{The flow time history of the topological charge (upper panels) and the gradient flow coupling (lower panels) on six selected gauge field configurations of our ten flavor $(L/a)^4$ ensembles with $L/a=32$, $L_s=16$ at bare coupling $\beta\equiv 6/g_0^2=4.02$. We show both Wilson (green) and Symanzik (red) flow and determine the gradient flow coupling for the Wilson-plaquette (W, solid lines) and clover (C, dashed lines) operator.   }
\label{fig:Nf10FlowHistory}
\end{figure*}

\begin{figure*}[tb]
  \includegraphics[width=0.95\columnwidth]{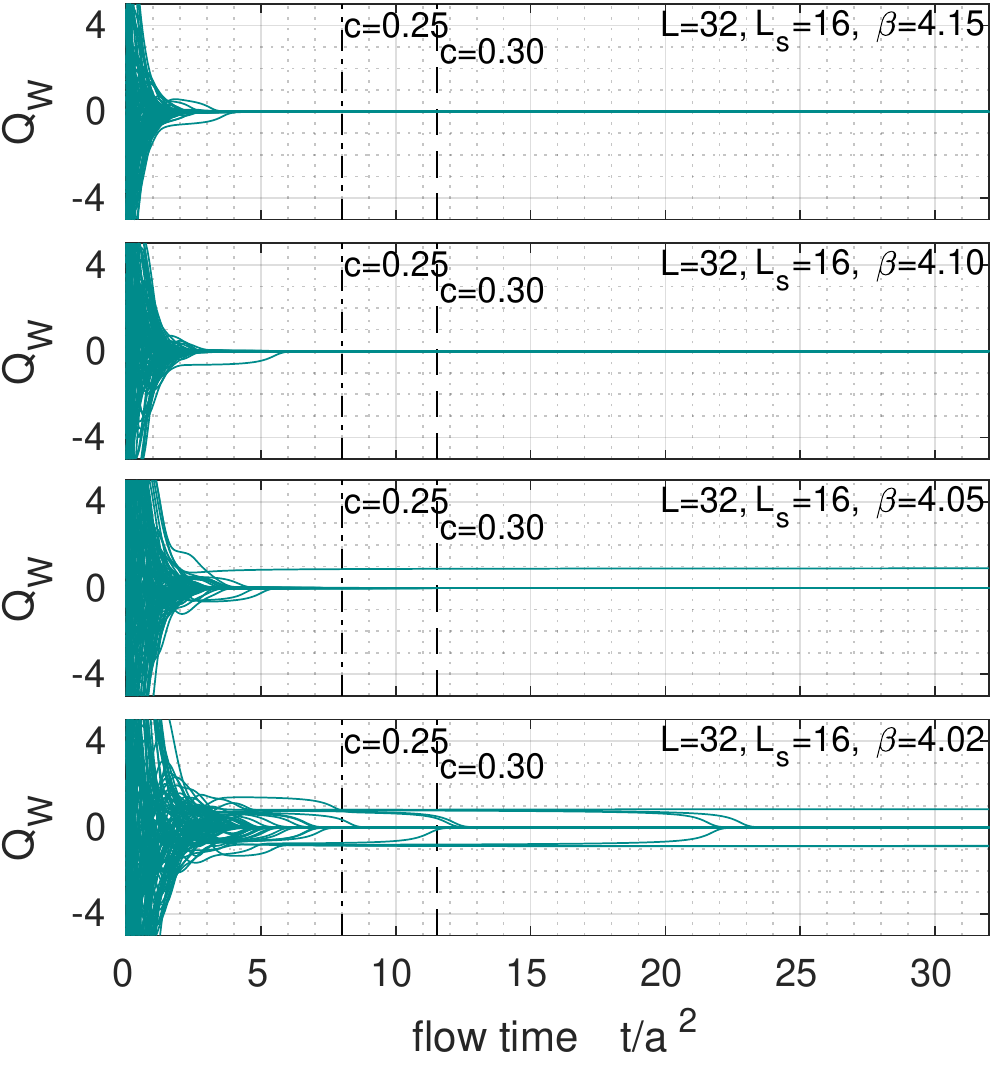}\hfill
  \includegraphics[width=0.95\columnwidth]{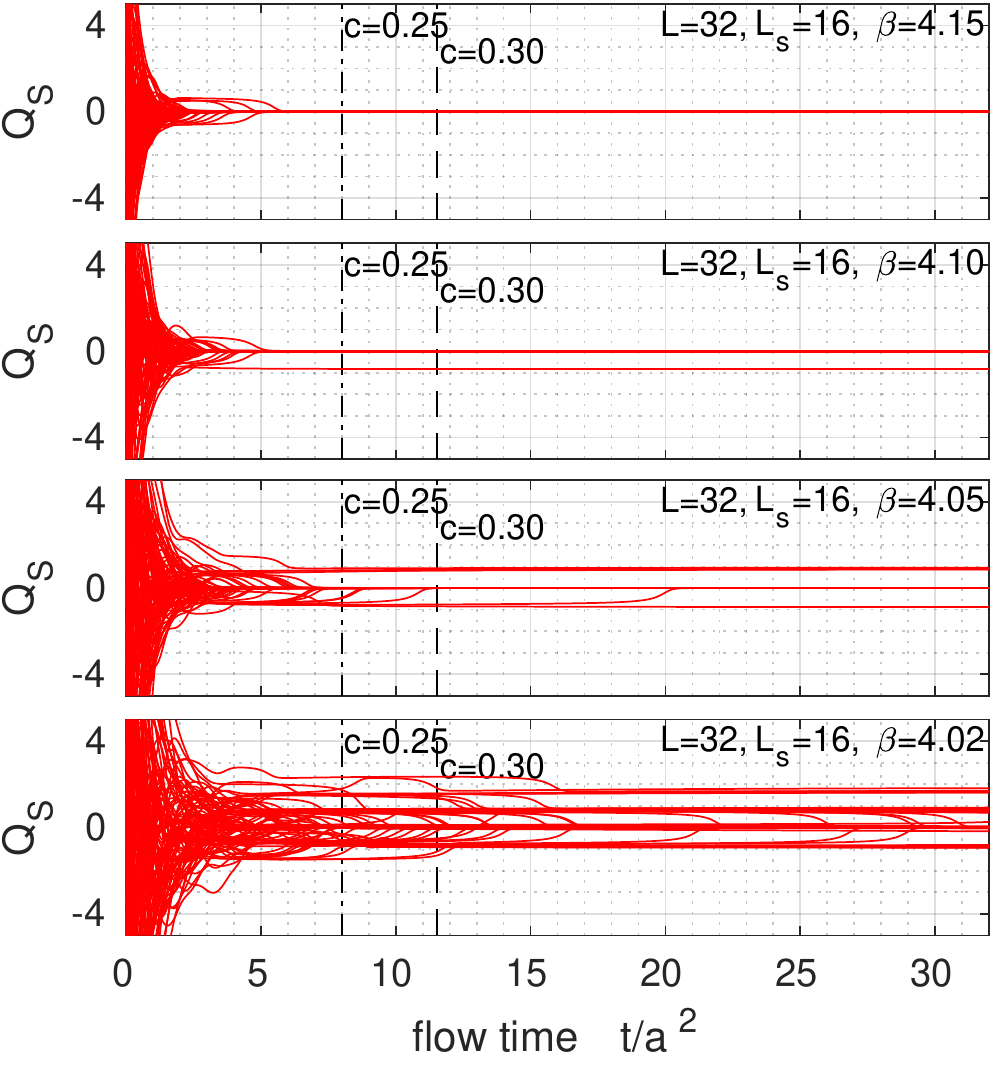}\hfill
  \caption{Dependence of the topological charge $Q$ on the flow time $t/a^2$ for $(L/a)^4=32^4$ ensemble with $L_s=16$ and at bare couplings $\beta=4.15$, 4.10, 4.05, and 4.02. Each panel show the flow time histories for the first thermalized 100 configurations of each ensemble. The left (right) panels shows the flow time histories using Wilson (Symanzik) gradient flow. Lattice artifacts in the form of nonzero topological charges $Q$ increase in the strong coupling limit (decreasing $\beta$) and are more pronounced for Symanzik than for Wilson flow. }
\label{fig:Nf10TopoHistory}
\end{figure*}

We illustrate the effects of $Q\ne 0$ instanton-like objects on the gradient flow coupling in Fig.~\ref{fig:Nf10FlowHistory}  where we show the flow time dependence of the topological charge $Q$ and the GF coupling $g^2_{GF}$ on six individual configurations. We use the clover operator to approximate $F \widetilde{F}$ in Eq.~(\ref{eq:Q_geom}). The upper panels in each sub-figure show the flow time evolution of the topological charge both with Wilson (W) and Symanzik (S) flows. The lower panels show the  the renormalized $g^2_{GF}$ coupling evaluated with both the Wilson plaquette (W) and clover  (C) operators for both flows.\footnote{The first letter shorthand notation indicates the gradient flow (W or S), the second letter the operator (W or C).}  The six configurations were chosen to illustrate the difference between $Q=0$ and $Q\ne 0$. They are  part of our $N_f=10$ DW ensemble  at  $\beta=4.02$, the strongest bare coupling we consider, on $32^4$ volumes \cite{Nf10stepScaling}.  The topological charge shows large fluctuations at small flow time but settles to a near-integer value by $t/a^2\gtrsim 5.0$.  We  observe occasional change in $Q$  for $t/a^2>5$ but  these tend to be quick as topological objects are annihilated by the flow. 

At large flow time we expect different flows and operators to converge. That is indeed the case at trajectory \#700 and \#575 (top left and top right in Fig.~\ref{fig:Nf10FlowHistory})   where, as shown in the upper panels,  Wilson and Symanzik flows find the same topological charge at large flow time.  Both Wilson and Symanzik flows and  Wilson and clover operators predict consistent   $g^2_{GF}$ at large flow time, as is shown on the lower panels. 

 At trajectory \#2965 and \#2100 (middle of Fig.~\ref{fig:Nf10FlowHistory})  Wilson flow predicts $Q=0$ but Symanzik flow identifies topological charge $Q=2$ and -2, respectively. With Wilson flow,  $g^2_{GF}$ shows a flat, slowly decreasing behavior with  flow time, similar to what is observed at trajectory \# 700 with $Q=0$.  Symanzik flow, however, shows $g^2_{GF}$ increasing roughly linearly with the flow time, similar to trajectory \# 575, $Q=-1$,  although the slope is larger, consistent with two topological objects on the configurations. Different operators are still consistent within each flow.

 At trajectory \#2255 (bottom left) and  \#845 (bottom right)  the topological charge with Wilson flow is $Q =0$  but with Symanzik flow we see a rapid change at larger flow time. At  trajectory \#2255 this corresponds to $Q=-1 \to Q=0$ around $t\approx 26$. Correspondingly $g^2_{GF}$ changes from a linearly increasing flow time dependence to a flat/decreasing form. At  trajectory  \#845 the change is $Q=0 \to Q=-1$, suggesting that the configuration at flow time $t/a^2 < 12$ had an instanton-antiinstanton pair. The instanton is annihilated by the flow at $t/a^2\approx 13$, leaving the anti-instanton unpaired.  The renormalized coupling $g^2_{GF}$ follows the expected behavior. Its linear rise with the flow time slows at $t/a^2\approx 13$ but remains linear, similar to what is observed at trajectory \#575.

 \begin{figure*}[tb]
  \includegraphics[width=0.95\columnwidth]{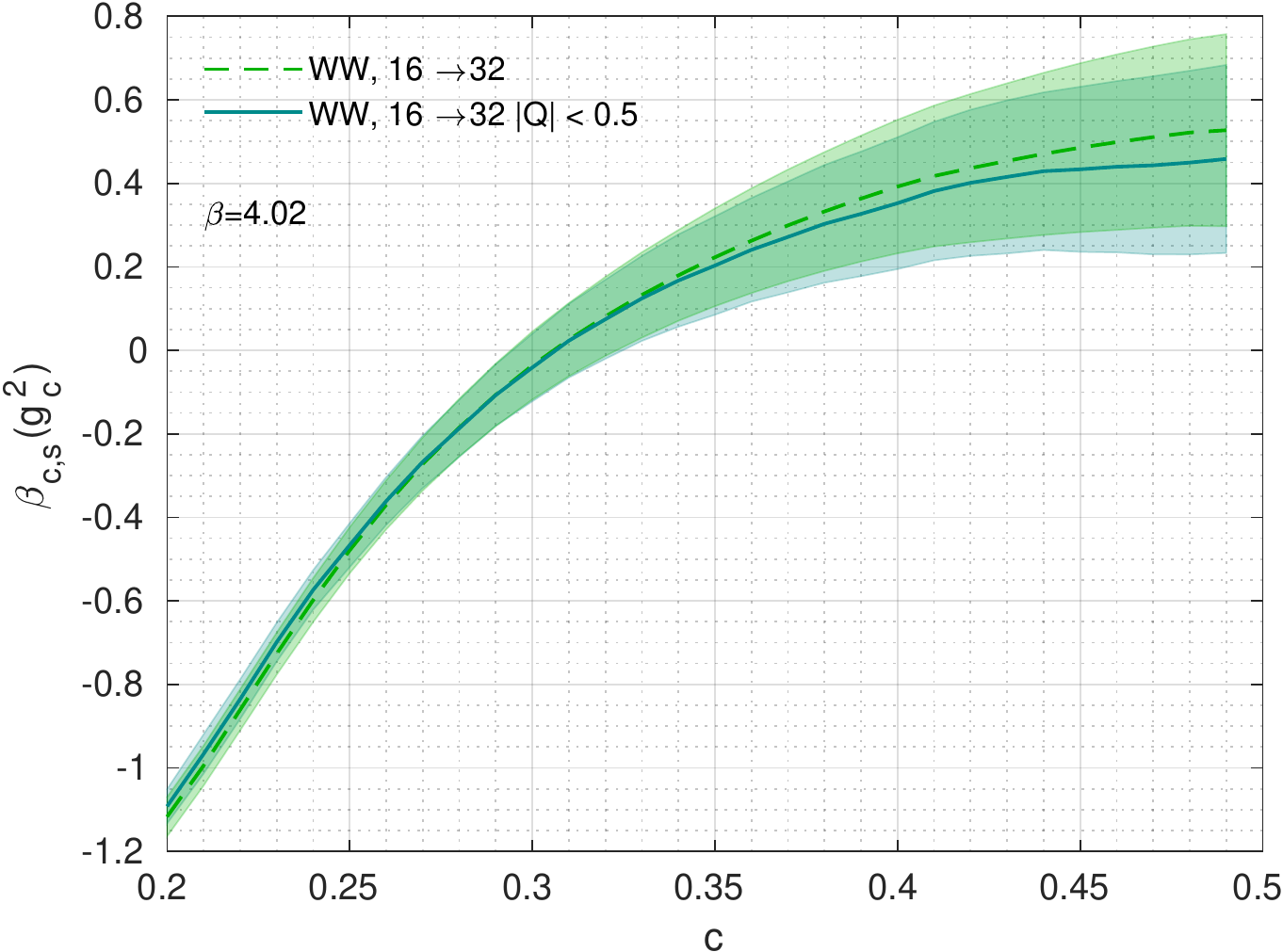}\hfill    
  \includegraphics[width=0.95\columnwidth]{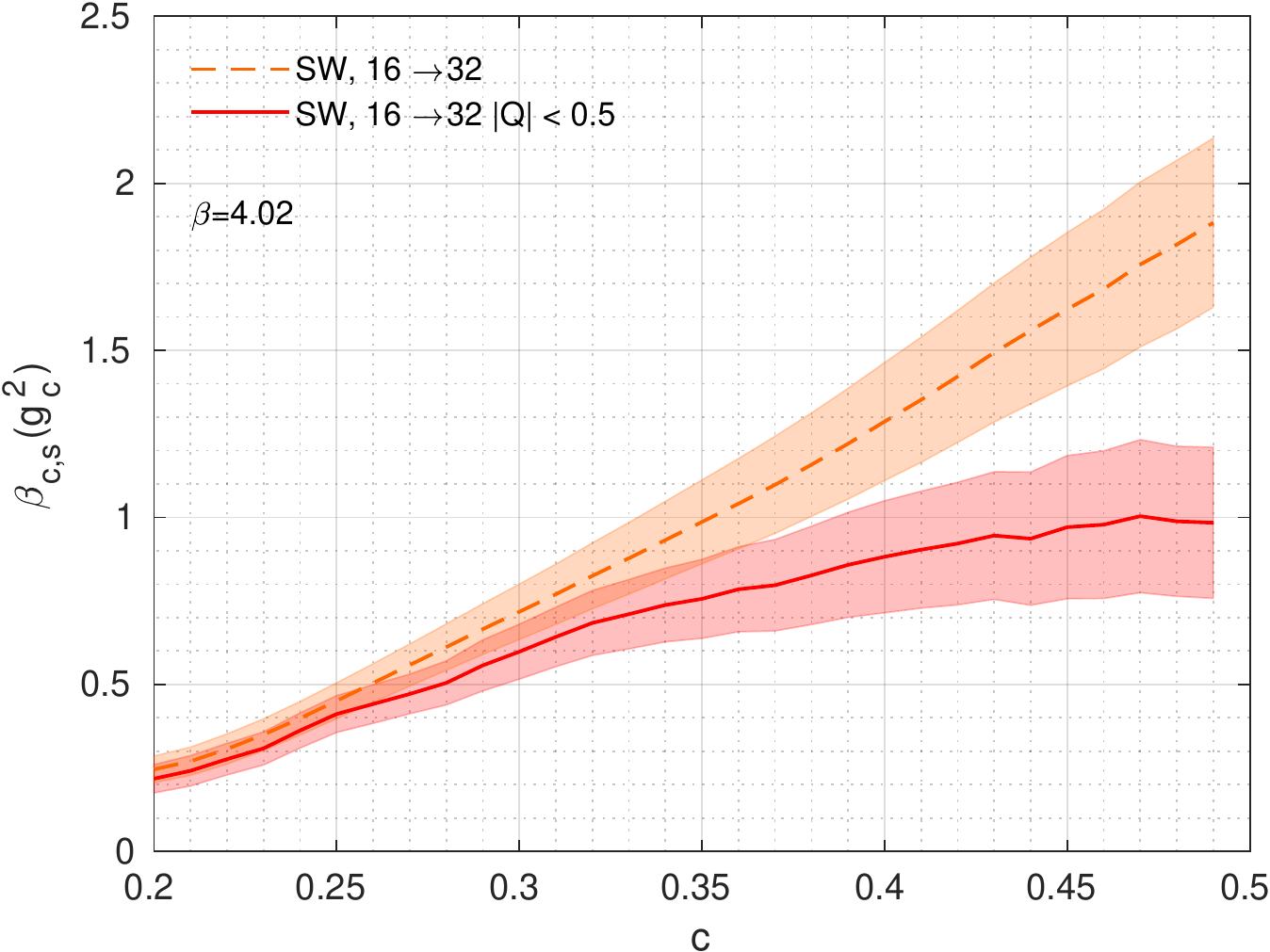}
\caption{Effect of non-zero topological charge $Q$ on the value of the step-scaling $\beta_{c,s}$ function determined from the Wilson plaquette operator for the $16 \to 32$ volume pair $(s=2)$ at bare coupling $\beta\equiv 6/g_0^2=4.02$ as function of the renormalization scheme parameter $c$, which is related to the flow time $t$. Since at $\beta=4.02$ Wilson flow exhibits in total only three configurations where a nonzero topological charge is measured, filtering configurations with $|Q|<0.5$ does not impact the prediction of $\beta_{c,s}$ (left panel). For Symanzik flow (right panel), however, a large number of configurations with nonzero topological charge are found, resulting in a larger $\beta_{c,s}$ compared to the analysis using only $|Q|<0.5$. The discrepancy grows with $c$.}
\label{fig:Nf10TopoBeta}
\end{figure*}

Any non-vanishing $Q$ is an artifact of the gradient flow in simulations with massless chirally symmetric fermions.  The panels of  Fig.~\ref{fig:Nf10FlowHistory}  verify that on $Q\ne 0$ configurations $g^2_{GF}$ receives a contribution that increases approximately linearly with flow time and  $|Q|$. 
Next  we investigate what fraction of  the configuration ensembles is affected by this lattice artifact.  In Fig.~\ref{fig:Nf10TopoHistory} we show the flow time evolution of the topological charge defined by Eq.~(\ref{eq:Q_geom}) on a subset of our $N_f=10$, $32^4$ configurations at bare coupling $\beta=4.02$, 4.05, 4.10 and 4.15. Each panel includes 100 configurations, separated by 10 MDTU,  analyzing Wilson flow data on the left,  Symanzik flow data on the right. The vertical lines indicate flow times $t/a^2=8.0$ and 11.52  which correspond to $c=0.25$ and 0.3 in step-scaling studies.

 At small flow time, vacuum fluctuations dominate $Q$.  At the weaker couplings, $\beta=4.15$,  4.10, most vacuum fluctuations die out by $t/a^2\gtrsim 2$, and  while $Q$  may not exactly be integer, it is close to an integer (0, $\pm 1$, $\pm 2$, \dots). It is well known that different gauge actions suppress/promote dislocations differently \cite{DeGrand:2000gq,DeGrand:2002vu}. The gradient flow has a similar dependence of the flow action. With Wilson flow most configurations have $Q=0$. Symanzik flow sustains $Q\ne0$ longer, and one of the 100 $\beta=4.10$  configurations remains $Q=-1$ even at $t/a^2=32$, our maximal flow time.
 The picture changes rapidly towards strong coupling. At our strongest gauge coupling $\beta = 4.02$ even Wilson flow has several $Q\ne0$ configurations at $t/a^2\approx 10$, some surviving even at $t/a^2=32$. Symanzik flow enhances this lattice artifact even further. The topological charge distribution of Symanzik flowed configurations resemble QCD at finite mass. Many QCD simulations would be relieved to see such a rapid change of the topological charge,  yet here $Q\ne 0$ signals only lattice artifacts.

At large flow time it is possible to filter configurations according to different topological sectors. Analyzing only those with $Q=0$ and contrasting the predictions with the full data set provides information on the effect of $Q\ne 0$. In Fig.~\ref{fig:Nf10TopoBeta} we compare the finite volume step-scaling $\beta_{c,s}(g^2_c)$ functions defined in Eq.~(\ref{eq:gGF_Q}) with and without topological filtering. The plots show  $\beta_{c,s}$ predicted by the lattice  volumes $L/a=16 \to 32$ on the $\beta=4.02$ configuration set as the function of $c=\sqrt{8 t}/L$ with Wilson flow (left) and Symanzik flow (right), using the Wilson plaquette operator to predict the energy density.  While in the Wilson flow analysis filtering on the topology has only a minimal effect,  the $Q=0$ subset with Symanzik flow predicts a significantly slower running step-scaling function\footnote{We define the integer topological charge as  the integer part of $(|Q_\text{geom}|+0.5)$ where $Q_\text{geom}$ is the value predicted by the clover $F{\tilde F}$ operator. At large flow time $Q_\text{geom}$ is close to an integer, apart from the regions where the topological charge undergoes a rapid change. }. This is consistent with  the observation we made in connection with Fig.~\ref{fig:Nf10FlowHistory} where we pointed out that $Q\ne 0$ configurations have faster running gauge coupling $g^2_{GF}$. This effect weakens at weak gauge coupling, but we expect that step-scaling functions could overestimate the running of the gauge coupling in the strong coupling, especially with Symanzik flow. In the accompanying paper \cite{Nf10stepScaling} we show details of our analysis.

\begin{figure}[tb]
\centering
\includegraphics[width=0.95\columnwidth]{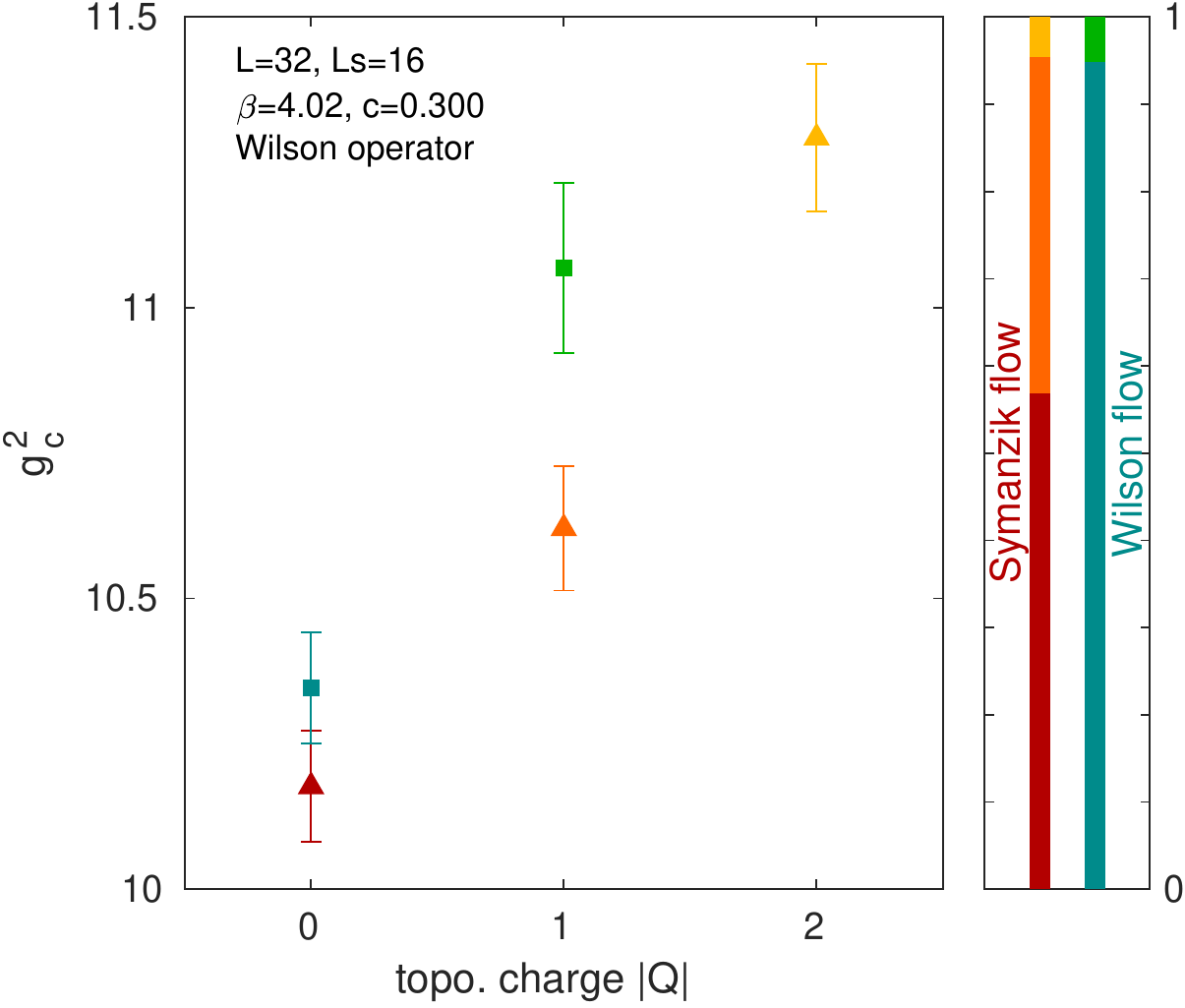} 
\caption{On the left: renormalized coupling $g^2_c$ as predicted by Symanzik (red/orange/yellow triangles) and Wilson (green squares) flows and Wilson operator on configurations with $|Q|=0$, 1, and 2 at $c=0.300$ (GF flow time $t=11.52$). The panel on the right shows the relative fraction of configurations at each $|Q|$ sector when using Symanzik and Wilson, respectively. In total 371 (372) configurations enter the presented results for Symanzik (Wilson) flow.}
\label{fig:Nf10TopogcSq}
\end{figure}

We close our discussion with Fig.~\ref{fig:Nf10TopogcSq} where we compare $g^2_c$ for c=0.300 as predicted by configurations with $|Q|=0$, 1 and 2 on our $\beta=4.02$ data set. As expected based on Eq.~(\ref{eq:gGF_Q}) and Figs.~\ref{fig:Nf10FlowHistory} and \ref{fig:Nf10TopoBeta}, $g^2_c$ increases with $|Q|$. On the right side panel of  Fig.~\ref{fig:Nf10TopogcSq}  we show the relative weight of the different topological sectors. In the case of Symanzik flow we analyze 371 measurements in total, 211 with $|Q|= 0$, 143 with $|Q|=1$ and 17 with $|Q|=2$ but do not show one measurement with $|Q|>2$. In the case of Wilson flow we analyze 353 configurations with $|Q|=0$ and 19 with $|Q|=1$. Differences between WW and SW determinations of $g_c^2$ indicate cutoff effects which are only supposed to disappear after taking the continuum limit. 

Measuring the total $Q=n_+-n_-$ does not give information on possible instanton--antiinstanton pairs. However the change of the slopes of $g^2_{GF}$ observed in Fig.~\ref{fig:Nf10FlowHistory} suggests that most $Q\ne 0$ configurations have only isolated instantons and not many pairs. We want to strongly emphasize that our  analysis filtering on the topological charge is not an alternative method to predict the running coupling and the step-scaling function. We  solely use it to show the expected change  due to lattice artifacts created by  $Q\ne 0$ configurations.

\subsection{\texorpdfstring{Finite value of $L_s$}{Finite value of Ls}}
\begin{figure}[tb]
\includegraphics[width=0.95\columnwidth]{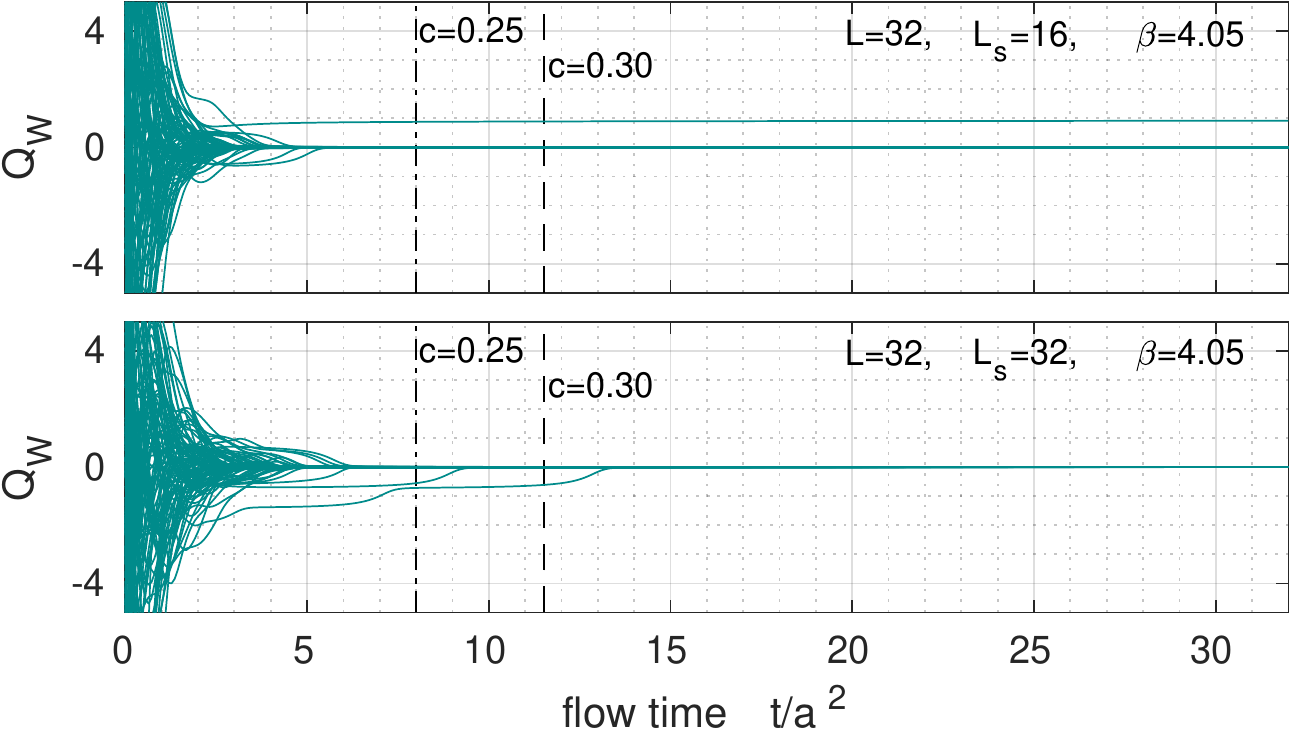}\\[5mm]
\includegraphics[width=0.95\columnwidth]{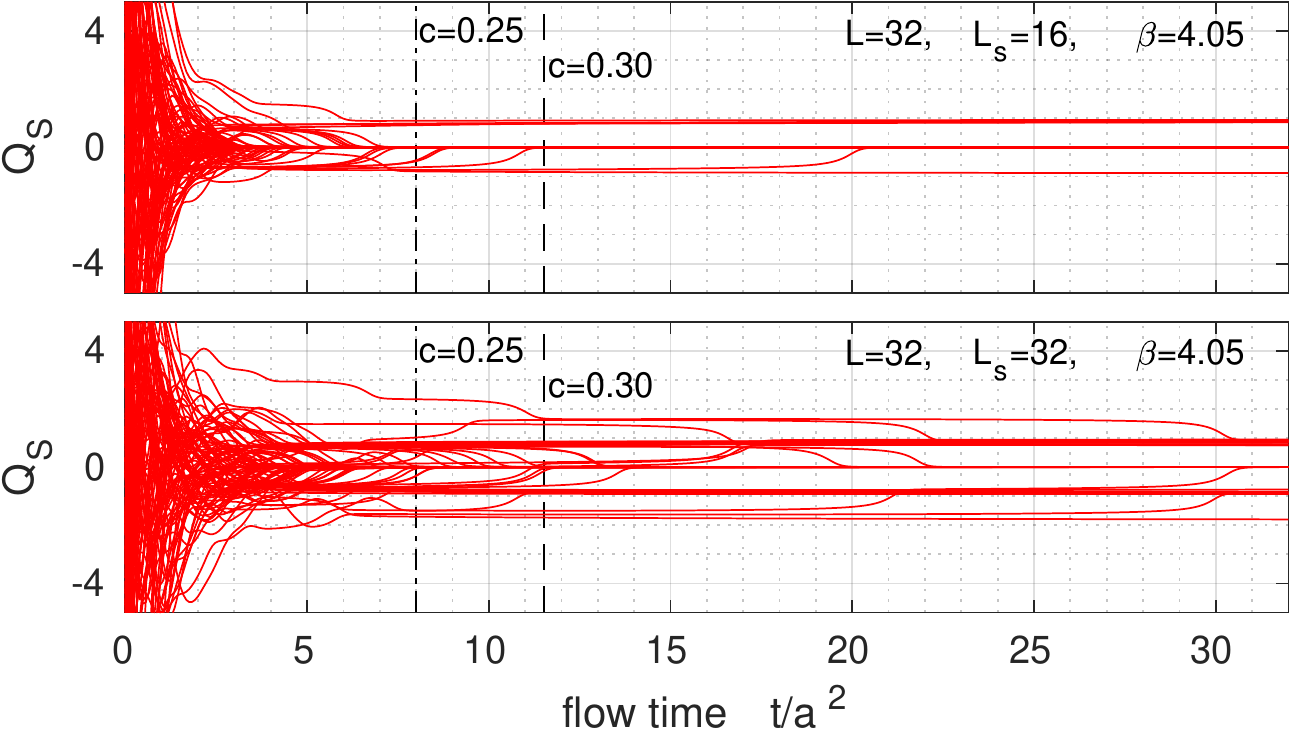}
\caption{ Comparison of flow time histories of the topological charge $Q$  for the first 100 thermalized configurations of the $\beta=4.05$ ensembles with $L_s=16$ and $L_s=32$. The upper two panels show $Q_W$ determined with Wilson flow (W), the lower two panels show $Q_S$ determined with Symanzik flow (S).  In each case, the $L_s=16$ data are shown above the $L_s=32$ data.}
\label{fig.Nf10TopoHistoryLs}  
\end{figure}

Stout smeared M\"obius domain wall fermions with $L_s=16$ have a small residual mass, $am_\text{res} < 10^{-3}$ even at our strongest gauge coupling. We check for possible effects due to non-vanishing residual mass by generating a second ensemble at bare coupling $\beta=4.05$ with $L_s=32$. The numerical cost of generating   an  $L_s=32$ trajectory is more than five times greater compared to the simulation with $L_s=16$. Thus we have fewer $L_s=32$ trajectories (about $1/3$) than for $L_s=16$. In Fig.~\ref{fig.Nf10TopoHistoryLs} we show the flow time histories for the topological charge $Q$ for the first 100 configurations of each ensemble. While Wilson flow identifies very few configurations with nonzero $Q$ on either ensembles, the same ensembles exhibit more nonzero topology under Symanzik flow.  Surprisingly, the relative number of configurations with nonzero $Q$ more than triples under Symanzik flow when $L_s$ increases from 16 to 32. This observation again indicates that  non-vanishing topology is  an artifact of the flow and not due to the small residual mass.

\begin{figure}[tb]
  \includegraphics[width=0.95\columnwidth]{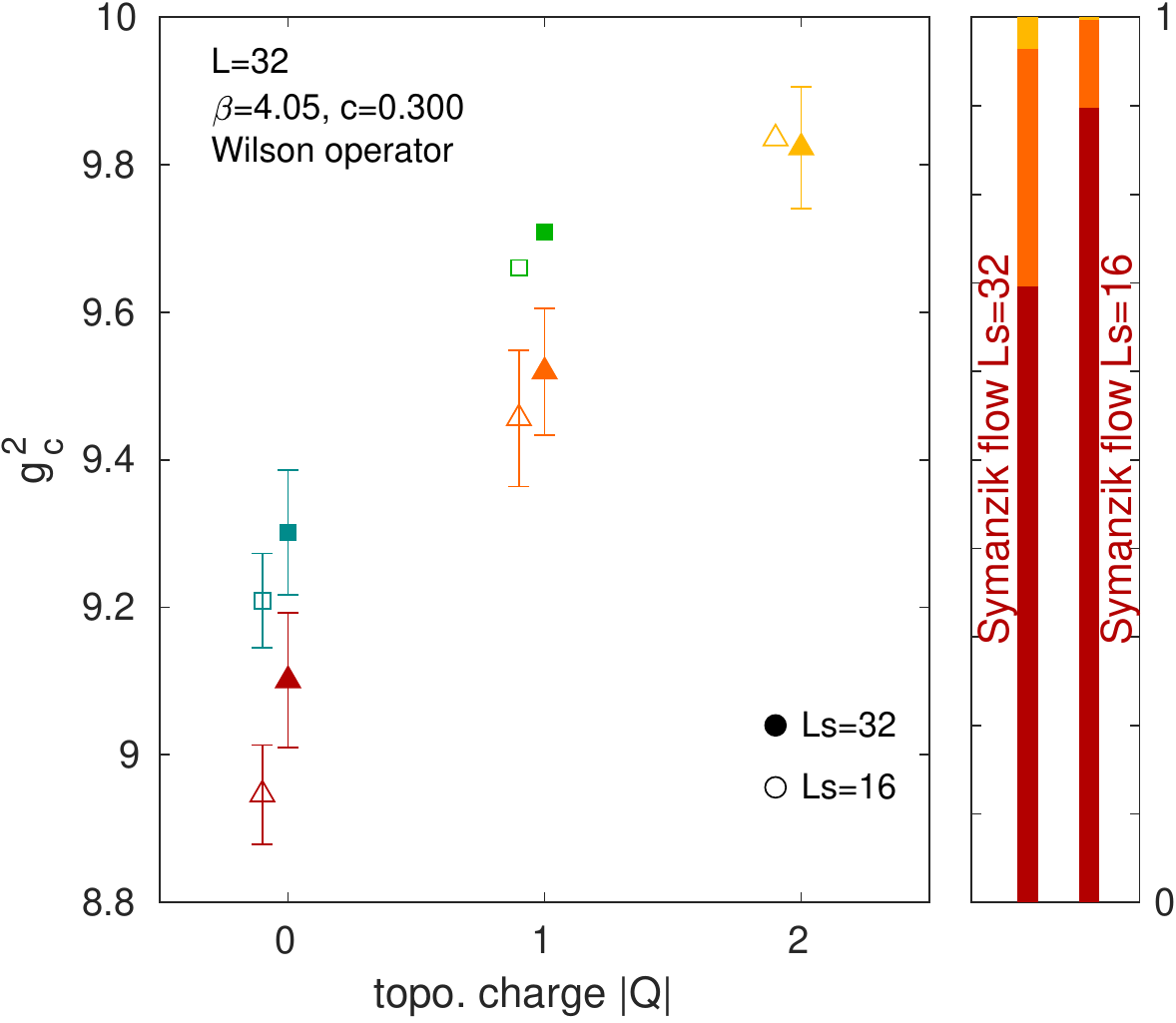}
  \caption{Renormalized coupling $g_c^2$ for the renormalization scheme $c=0.30$ determined on $L/a=32$ ensembles at $\beta=4.05$ for ensembles with $L_s=16$ (open symbols) and $L_s=32$ (filled symbols). In case of Wilson flow (green squares) only a value for configurations with $|Q|=0$ can be determined, whereas for Symanzik flow (red/orange/yellow triangles) $g_c^2$ for $|Q|=0,\,1$ sectors can be determined at $L_s=16$ and in addition $|Q|=2$ for $L_s=32$. All $g_c^2$ values at the same $|Q|$ agree perfectly which strongly implies effects due to $L_s=16$ are not resolved within our statistical errors. For Symanzik flow, however, each $|Q|$ sector predicts a statistically different value of $g_c^2$. The relative distribution of the $|Q|$ sectors for Symanzik flow are shown in the small panel on the right. For $L_s=16$ in total 372 measurements are analyzed, for $L_s=32$ 112.}
\label{fig.Nf10TopogcSqLs}  
\end{figure}

Next we determine the renormalized coupling $g_c^2$ for the renormalization scheme $c=0.300$ on both ensembles where we again separate configurations according to the value of $|Q|$. The outcome is shown in Fig.~\ref{fig.Nf10TopogcSqLs}. On both  ensembles Wilson flow (green squares) predominantly finds zero topological charge and identifies too few configurations with $|Q|=1$ to reliably estimate an uncertainty on $g_c^2$. Hence we  show the values for $|Q|=0$ with statistical error bar but indicate only the central values at $|Q|=1$.  The prediction on the $L_s=16$ ensemble (open symbol) is in perfect agreement with the $L_s=32$ ensemble (filled symbol). For Symanzik flow (red/orange/yellow triangles) we find several configurations with $|Q|=0$ and 1 plus one configuration with $|Q|=2$ on the $L_s=16$ ensemble and several configurations in all three sectors for $L_s=32$. The $g_c^2$ values clearly resolve a dependence on $Q$.  At the same time,  we observe good agreement for $g_c^2$ predicted at the same value of $Q$ on ensembles with different $L_s$. The latter strongly implies that the effect of choosing $L_s=16$ vs.~$L_s=32$ is negligible within our statistical uncertainties. The relative distribution of the $|Q|$ sectors for Symanzik flow are shown in the small panel on the right of Fig.~\ref{fig.Nf10TopogcSqLs}.  For $L_s=16$ in total 372 measurements are analyzed and 90\% have $Q=0$. For $L_s=32$ we analyze 112 measurements  but only 70\% have $|Q|=0$. Since $|Q|>0$ predict larger $g^2_c$, the average of the renormalized coupling increases with increasing $L_s$. However, this is an artifact of the flow and implies larger lattice artifacts for larger $L_s$.  

\section{\texorpdfstring{SU(3) with $N_f=8$ flavors}{SU(3) with Nf=8 flavors}}
\label{Sec.Nf8}
\subsection{Details of the simulations}

In this part of our study we utilize existing gauge field configurations generated with eight degenerate and massless flavors of staggered fermions with nHYP smeared links~\cite{Hasenfratz:2001hp,Hasenfratz:2007rf} and gauge action that combines plaquette and adjoint plaquette terms \cite{Hasenfratz:2014rna}. The configurations have symmetric volumes, $V=L^4$, where the gauge fields have periodic boundary conditions and the fermions antiperiodic boundary conditions in all four space-time directions. Apart from the boundary conditions this is the same action used in the large scale studies of Refs.~\cite{Appelquist:2016viq,Appelquist:2018yqe}. 

Staggered fermions have a remnant U(1) chiral symmetry that protects the fermion mass from additive mass renormalization. On the other hand taste breaking of staggered fermions split the eigenmodes of the Dirac operator. Smooth, isolated instantons have four near-zero  eigenmodes for the four staggered species, but they are split into two positive, two negative imaginary eigenvalue pairs. The determinant of the Dirac operator is not exactly zero, topologically non-trivial configurations are not prohibited. In the continuum limit taste symmetry is recovered and $Q\ne 0$ configurations should be suppressed. Therefore it is reasonable to consider all $Q\ne 0$ as lattice artifact --- either from the action or from the flow.

\subsection{Effects of nonzero topological charge}

\begin{figure*}[tb]
  \includegraphics[width=0.95\columnwidth]{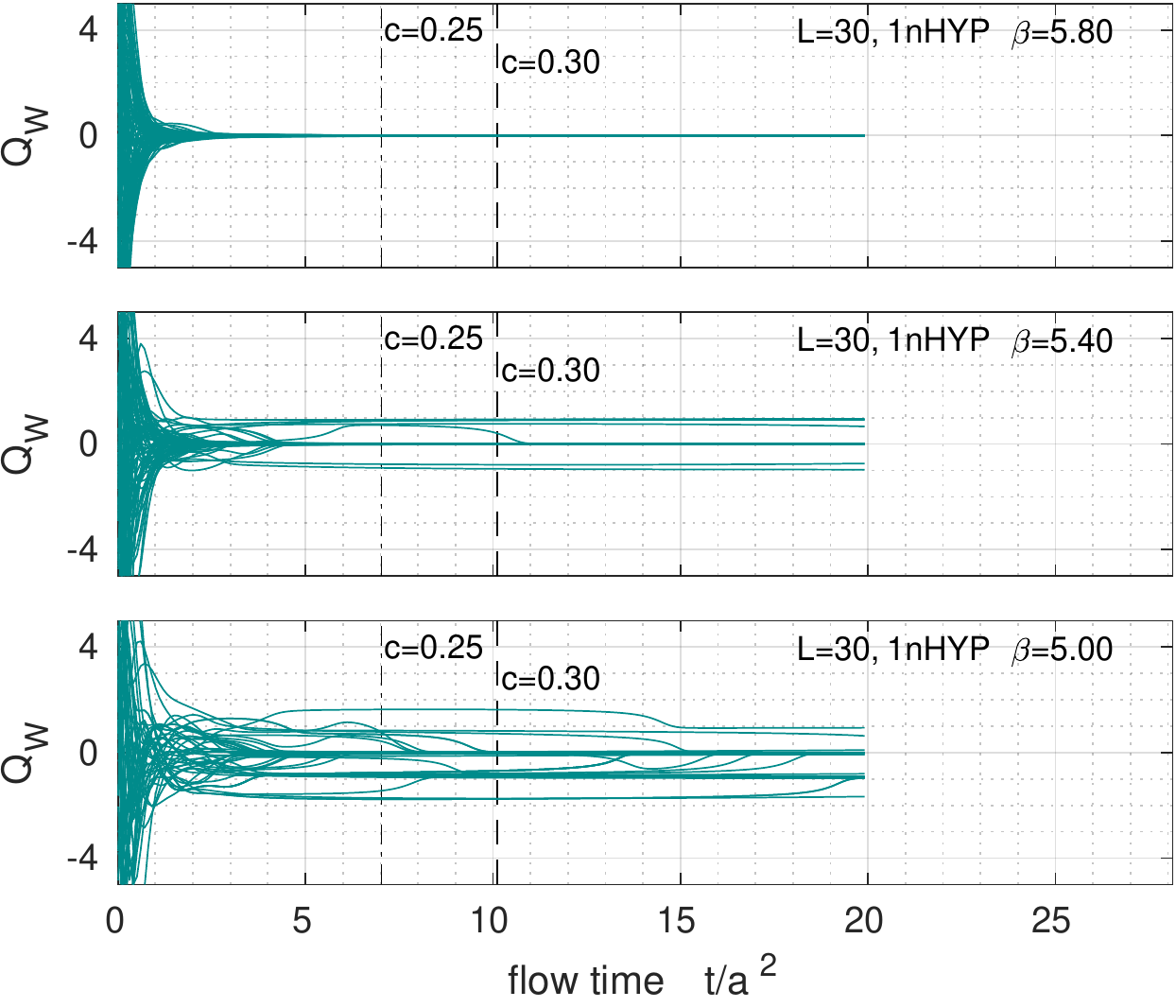}\hfill
  \includegraphics[width=0.95\columnwidth]{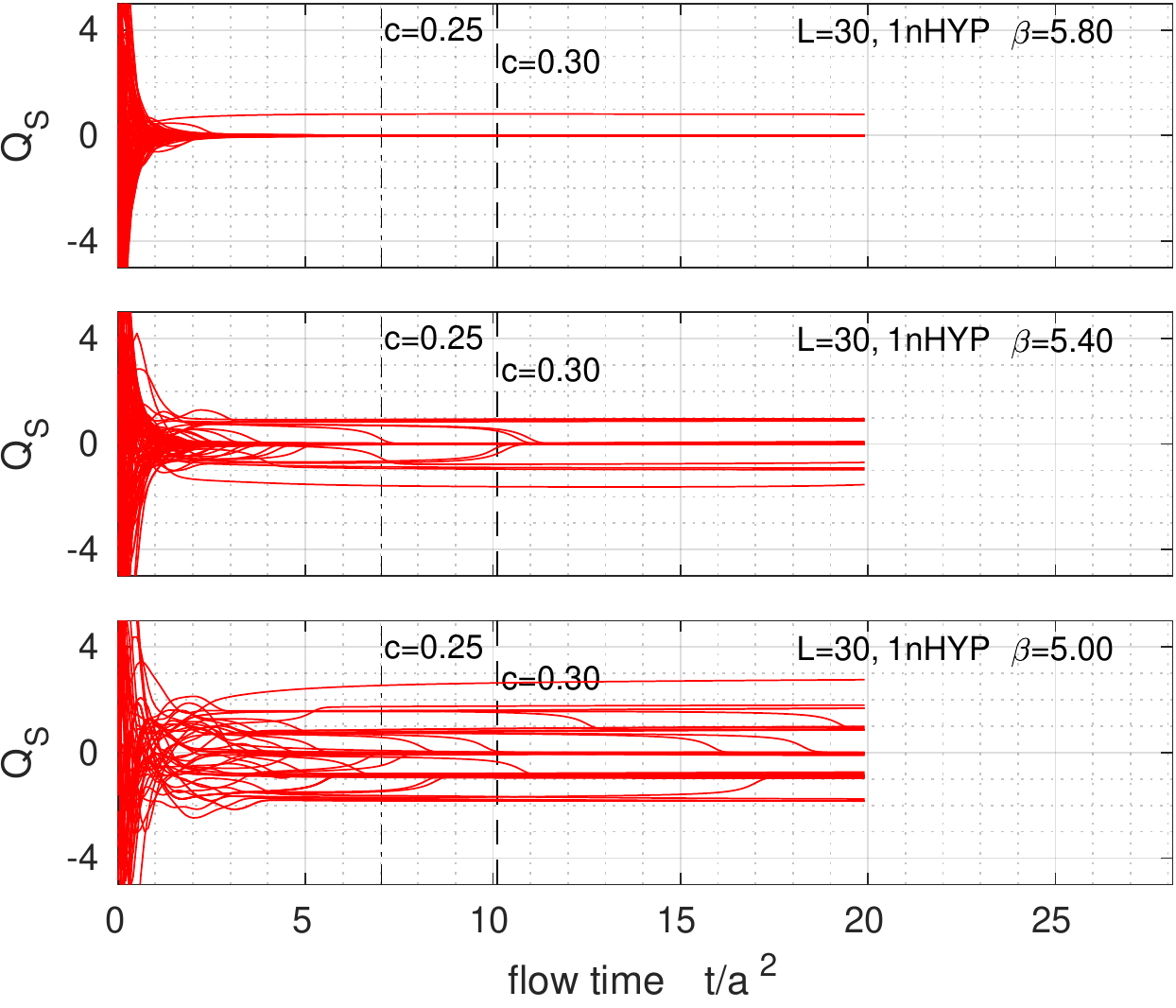}
  \caption{Flow time histories for the $N_f=8$ data set with one level of nHYP smeared staggered fermions on $(L/a)^4 =30^4$ lattices at bare gauge coupling $\beta=5.80,\, 5.40,\, 5.00$. Similar to $N_f=10$ DW fermions (Fig.~\protect{\ref{fig:Nf10FlowHistory})} we observe an increase of nonzero topology as $\beta$ decreases and the suppression of topology is inferior for Symanzik compared to Wilson flow.}
\label{fig:Nf8TopoHistory_1nHYP}  
\end{figure*}

\begin{figure}[tb]
  \includegraphics[width=0.95\columnwidth]{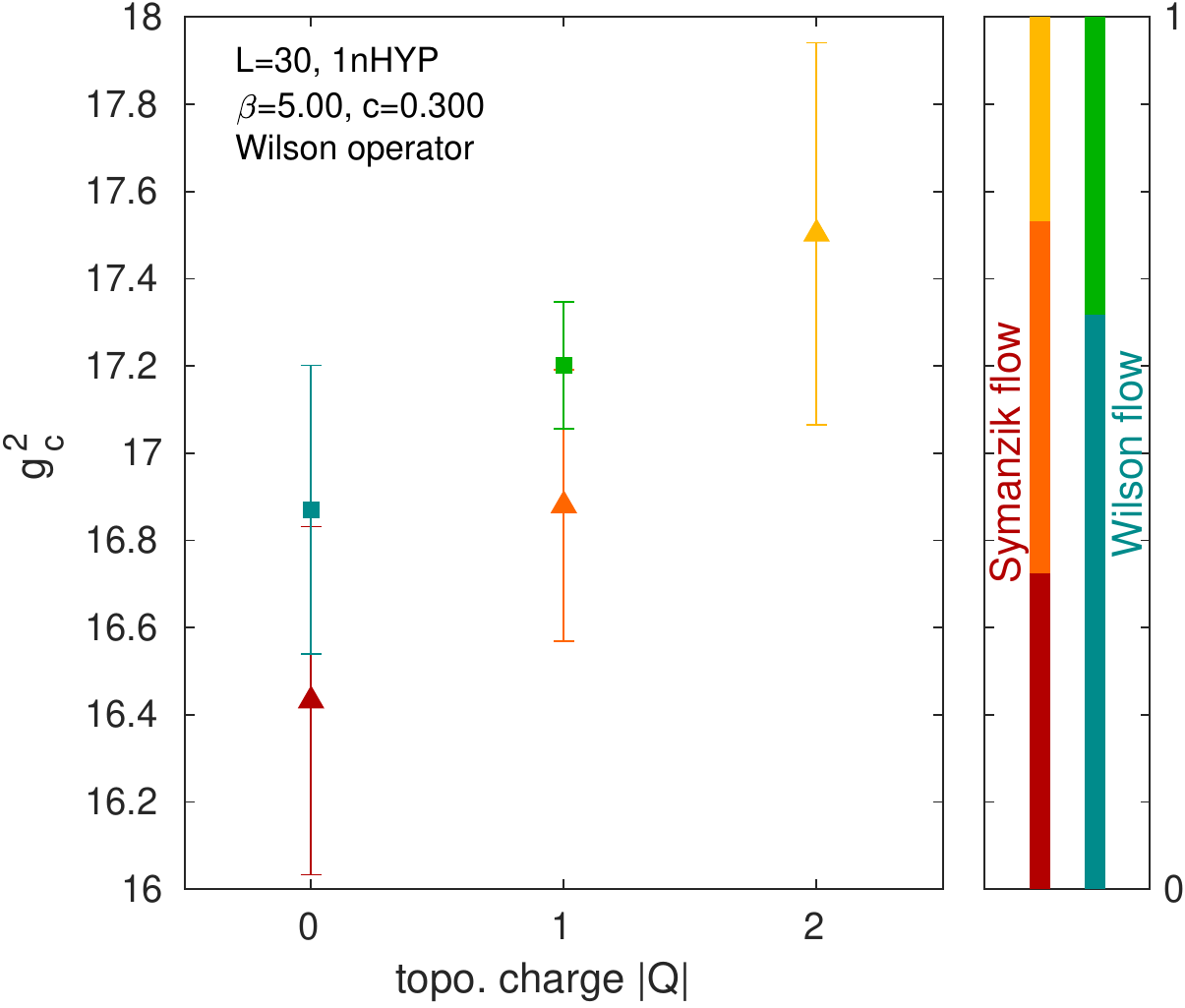}\\
 \caption{Renormalized coupling $g_c^2$ determined for different topological sectors using the strongest coupling of the $N_f=8$ staggered fermion ensemble at $\beta=5.00$ and $c=0.300$. $g_c^2$ increases with the charge $Q$, similar to DW in Fig.~\ref{fig:Nf10TopogcSq}.  Green squares correspond to Wilson flow, red/orange/yellow triangles to Symanzik flow. The relative fraction of each $Q$ sector is shown by the panel on the right. Similar to the DW results, Symanzik flow creates more $Q\ne 0$ configurations than Wilson flow.}
\label{fig:Nf8TopogcSq_2nHYP}  
\end{figure}

Our discussion and analysis here follows that of Sec.~\ref{Sec.Nf10} with domain wall fermions. The strongest gauge coupling of the simulations with one level of nHYP smearing is $\beta=5.0$, and the largest volume has $L/a=30$. In Fig.~\ref{fig:Nf8TopoHistory_1nHYP} we show the evolution of the topological charge with Wilson and Symanzik flow on 50 thermalized consecutive  configurations at $\beta=5.0$, 5.4 and 5.8. Similar to the DW result, we observe the emergence of  more  $Q\ne 0$ configurations at strong coupling. We also observe rapid changes in $Q$ at large flow time, and again more $|Q|>0$ with Symanzik than with Wilson flow. In Fig.~\ref{fig:Nf8TopogcSq_2nHYP}  we compare the renormalized GF coupling in the  $c=0.300$ renormalization scheme  for the different topological sectors. As in Fig.~\ref{fig:Nf10TopogcSq}, we see a clear increase in $g^2_c$ as $|Q|$ increases. Since the fraction of $Q\ne 0$ configurations is much larger with Symanzik than Wilson flow, this implies that step scaling studies using Symanzik flow may overestimate $\beta_{c,s}(g^2)$ at strong gauge coupling.

We note however the investigation in  Ref.~\cite{Hasenfratz:2001hp}  studied this system  using only Wilson flow. It would be very interesting to re-analyze the existing configurations not only with Symanzik flow, but also with a flow that suppresses the topology even further that Wilson flow. 

\section{Gradient flow with improved  topology suppression}
\label{Sec.Tflow}
\begin{figure}[tb]
\includegraphics[width=0.95\columnwidth]{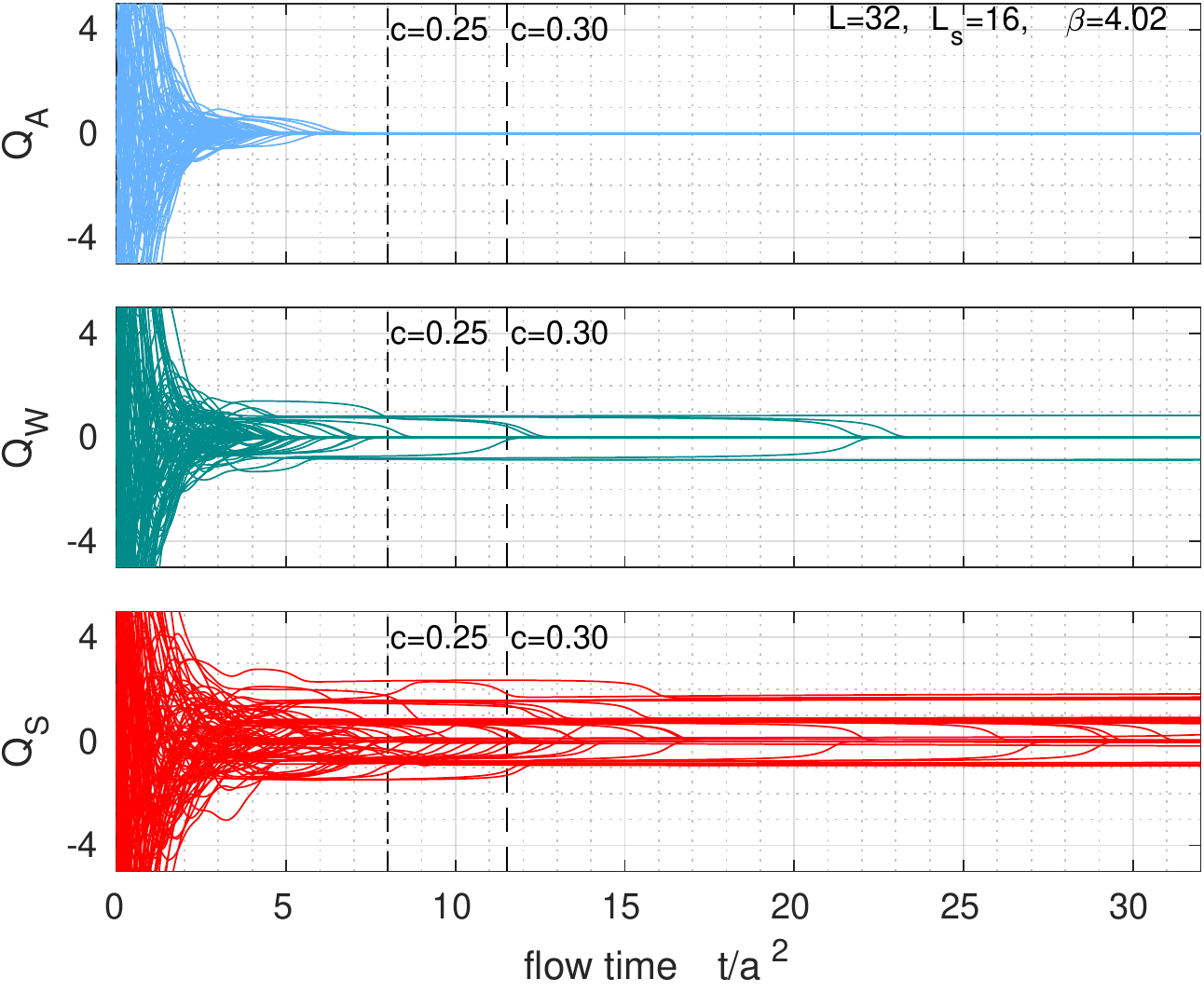}
\caption{Demonstration of the topology suppressing features of our alternative flow with positive coefficient for the $2\times 1$ rectangle term. We explore this alternative GF using our $N_f=10$ domain wall lattice with $L/a=32$ at $\beta=4.02$ and analyze again the first 100 thermalized configurations. The alternative flow (A) is shown on the top in blue and for easy comparison we repeat below Wilson flow (W, green) and Symanzik flow (S, red).}
\label{fig.Tflow}  
\end{figure}

The flow kernel of Symanzik flow is a combination of a $1\times 1$ plaquette and a $2\times 1$ rectangle term, with coefficients $c_{1\times 1}=5/3$ and $c_{2\times 1} =-1/12$.  Wilson flow is performed only with the plaquette term i.e.~$c_{1\times 1}=1$, $c_{2\times 1} =0$. Apparently the negative $c_{2\times 1}$  term increases the probability of $Q\ne 0$ in Symanzik flow. This suggests that a positive $c_{2\times 1}$ term might lead to  a better suppression of this lattice artifact. To test the idea we implemented an alternative gradient flow (A) where we set the coefficients to
\begin{align}
  c_{1\times 1} =2/3 \quad \text{and}\quad  c_{2\times 1} =1/24
\end{align}  
and demonstrate its effect on the topological charge $Q$ using our $N_f=10$ domain wall ensemble at bare coupling $\beta=4.02$. In Fig.~\ref{fig.Tflow} we show how the suppression of the topological charge is improved w.r.t.~Wilson and Symanzik flow. Whether or not this alternative gradient flow is a viable candidate to perform step-scaling studies at strong coupling will however require further  investigations using multiple volumes and a range of bare coupling $\beta$. Only that will allow to estimate  discretization effects to be removed by the continuum limit extrapolation.

\section{Summary}
\label{Sec.summary}

In this paper we demonstrate that gradient flow measurements on rough gauge field configurations can promote lattice dislocations to instanton-like topological objects. 
The number of these instanton-like objects depend on the gradient flow kernel. In the case of step-scaling calculations of the lattice $\beta$ function, the simulations are carried out in the chiral limit where a nonzero instanton number is suppressed. Hence instanton-like objects created by the gradient flow are lattice artifacts. Our investigations reveal a clear correlation between a nonzero topological charge seen  by the gradient flow and an increase in the value of gradient flow renormalized coupling.   We further demonstrate that this also results in an overestimate of the step-scaling $\beta$-function. By investigating the $N_f=10$ system simulated with domain wall fermions and the $N_f=8$ system studied with staggered fermions, we show that this artifact is not related to the lattice actions used in the simulations but an artifact of the gradient flow which arises at (very) strong coupling.  In both systems we also observe that the effect is more pronounced when using Symanzik compared to Wilson flow.

Since this effect becomes only noticeable at very strong coupling, it may explain why it has not been reported earlier. In the case of our $N_f=12$ simulations, we checked that both step-scaling calculations  using domain wall \cite{Hasenfratz:2019dpr,Hasenfratz:2018wpq,Hasenfratz:2017qyr} or staggered fermions \cite{Hasenfratz:2016dou}  do not include ensembles exhibiting more than one or two configurations where a gradient flow finds nonzero topological charge. These simulations have simply been performed at weaker coupling. 

Similarly to step-scaling calculations,  continuous $\beta$ function  determinations \cite{Hasenfratz:2019hpg,Hasenfratz:2019puu,Fodor:2017die}  at (very) strong coupling might also be affected by nonzero topological charge occurring as part of the gradient flow. Our studies of the $N_f=2$ and 12 systems, however, do not extend into the problematic range and are therefore not affected.

\begin{acknowledgments}

  We are very grateful to Peter Boyle, Guido Cossu, Anontin Portelli, and Azusa Yamaguchi who develop the \texttt{Grid} software library providing the basis of this work and who assisted us in installing and running \texttt{Grid} on different architectures and computing centers. We thank David Schaich for his support to run additional Symanzik flow measurements on the $N_f = 8$ staggered configurations of Ref.~\cite{Hasenfratz:2014rna}.  We benefited from many  discussions with Thomas DeGrand, Ethan Neil,  and Benjamin Svetitsky.  A.H.~and O.W.~acknowledge support by DOE grant DE-SC0010005.
A.H.~would like to acknowledge the Mainz Institute for Theoretical Physics (MITP) of the Cluster of Excellence PRISMA+ (Project ID 39083149) for enabling us to complete a portion of this work. 
O.W.~acknowledges partial support by the Munich Institute for Astro- and Particle Physics (MIAPP) which is funded by the Deutsche Forschungsgemeinschaft (DFG, German Research Foundation) under Germany's Excellence Strategy -- EXC-2094 -- 390783311. 

Computations for this work were carried out in part on facilities of the USQCD Collaboration, which are funded by the Office of Science of the U.S.~Department of Energy and the RMACC Summit supercomputer \cite{UCsummit}, which is supported by the National Science Foundation (awards ACI-1532235 and ACI-1532236), the University of Colorado Boulder, and Colorado State University. This work used the Extreme Science and Engineering Discovery Environment (XSEDE), which is supported by National Science Foundation grant number ACI-1548562 \cite{xsede} through allocation TG-PHY180005 on the XSEDE resource \texttt{stampede2}.  This research also used resources of the National Energy Research Scientific Computing Center (NERSC), a U.S. Department of Energy Office of Science User Facility operated under Contract No. DE-AC02-05CH11231.  We thank  Fermilab,  Jefferson Lab, NERSC, the University of Colorado Boulder, TACC, the NSF, and the U.S.~DOE for providing the facilities essential for the completion of this work. 

\end{acknowledgments}

\bibliography{../General/BSM}

\begin{thebibliography}{61}%
\makeatletter
\providecommand \@ifxundefined [1]{%
 \@ifx{#1\undefined}
}%
\providecommand \@ifnum [1]{%
 \ifnum #1\expandafter \@firstoftwo
 \else \expandafter \@secondoftwo
 \fi
}%
\providecommand \@ifx [1]{%
 \ifx #1\expandafter \@firstoftwo
 \else \expandafter \@secondoftwo
 \fi
}%
\providecommand \natexlab [1]{#1}%
\providecommand \enquote  [1]{``#1''}%
\providecommand \bibnamefont  [1]{#1}%
\providecommand \bibfnamefont [1]{#1}%
\providecommand \citenamefont [1]{#1}%
\providecommand \href@noop [0]{\@secondoftwo}%
\providecommand \href [0]{\begingroup \@sanitize@url \@href}%
\providecommand \@href[1]{\@@startlink{#1}\@@href}%
\providecommand \@@href[1]{\endgroup#1\@@endlink}%
\providecommand \@sanitize@url [0]{\catcode `\\12\catcode `\$12\catcode
  `\&12\catcode `\#12\catcode `\^12\catcode `\_12\catcode `\%12\relax}%
\providecommand \@@startlink[1]{}%
\providecommand \@@endlink[0]{}%
\providecommand \url  [0]{\begingroup\@sanitize@url \@url }%
\providecommand \@url [1]{\endgroup\@href {#1}{\urlprefix }}%
\providecommand \urlprefix  [0]{URL }%
\providecommand \Eprint [0]{\href }%
\providecommand \doibase [0]{http://dx.doi.org/}%
\providecommand \selectlanguage [0]{\@gobble}%
\providecommand \bibinfo  [0]{\@secondoftwo}%
\providecommand \bibfield  [0]{\@secondoftwo}%
\providecommand \translation [1]{[#1]}%
\providecommand \BibitemOpen [0]{}%
\providecommand \bibitemStop [0]{}%
\providecommand \bibitemNoStop [0]{.\EOS\space}%
\providecommand \EOS [0]{\spacefactor3000\relax}%
\providecommand \BibitemShut  [1]{\csname bibitem#1\endcsname}%
\let\auto@bib@innerbib\@empty
\bibitem [{\citenamefont {Schäfer}\ and\ \citenamefont
  {Shuryak}(1998)}]{Schafer:1996wv}%
  \BibitemOpen
  \bibfield  {author} {\bibinfo {author} {\bibfnamefont {T.}~\bibnamefont
  {Schäfer}}\ and\ \bibinfo {author} {\bibfnamefont {E.~V.}\ \bibnamefont
  {Shuryak}},\ }\href {\doibase 10.1103/RevModPhys.70.323} {\bibfield
  {journal} {\bibinfo  {journal} {Rev. Mod. Phys.}\ }\textbf {\bibinfo {volume}
  {70}},\ \bibinfo {pages} {323} (\bibinfo {year} {1998})},\ \Eprint
  {http://arxiv.org/abs/hep-ph/9610451} {arXiv:hep-ph/9610451 [hep-ph]}
  \BibitemShut {NoStop}%
\bibitem [{\citenamefont {DeGrand}\ and\ \citenamefont
  {Hasenfratz}(2001)}]{DeGrand:2000gq}%
  \BibitemOpen
  \bibfield  {author} {\bibinfo {author} {\bibfnamefont {T.~A.}\ \bibnamefont
  {DeGrand}}\ and\ \bibinfo {author} {\bibfnamefont {A.}~\bibnamefont
  {Hasenfratz}},\ }\href {\doibase 10.1103/PhysRevD.64.034512} {\bibfield
  {journal} {\bibinfo  {journal} {Phys. Rev.}\ }\textbf {\bibinfo {volume}
  {D64}},\ \bibinfo {pages} {034512} (\bibinfo {year} {2001})},\ \Eprint
  {http://arxiv.org/abs/hep-lat/0012021} {arXiv:hep-lat/0012021 [hep-lat]}
  \BibitemShut {NoStop}%
\bibitem [{\citenamefont {DeGrand}\ \emph {et~al.}(2003)\citenamefont
  {DeGrand}, \citenamefont {Hasenfratz},\ and\ \citenamefont
  {Kovacs}}]{DeGrand:2002vu}%
  \BibitemOpen
  \bibfield  {author} {\bibinfo {author} {\bibfnamefont {T.~A.}\ \bibnamefont
  {DeGrand}}, \bibinfo {author} {\bibfnamefont {A.}~\bibnamefont {Hasenfratz}},
  \ and\ \bibinfo {author} {\bibfnamefont {T.~G.}\ \bibnamefont {Kovacs}},\
  }\href {\doibase 10.1103/PhysRevD.67.054501} {\bibfield  {journal} {\bibinfo
  {journal} {Phys. Rev.}\ }\textbf {\bibinfo {volume} {D67}},\ \bibinfo {pages}
  {054501} (\bibinfo {year} {2003})},\ \Eprint
  {http://arxiv.org/abs/hep-lat/0211006} {arXiv:hep-lat/0211006 [hep-lat]}
  \BibitemShut {NoStop}%
\bibitem [{\citenamefont {Bernard}\ \emph {et~al.}(2003)\citenamefont
  {Bernard}, \citenamefont {DeGrand}, \citenamefont {Hasenfratz}, \citenamefont
  {Detar}, \citenamefont {Osborn}, \citenamefont {Gottlieb}, \citenamefont
  {Gregory}, \citenamefont {Toussaint}, \citenamefont {Hart}, \citenamefont
  {Heller}, \citenamefont {Hetrick},\ and\ \citenamefont
  {Sugar}}]{Bernard:2003gq}%
  \BibitemOpen
  \bibfield  {author} {\bibinfo {author} {\bibfnamefont {C.}~\bibnamefont
  {Bernard}}, \bibinfo {author} {\bibfnamefont {T.~A.}\ \bibnamefont
  {DeGrand}}, \bibinfo {author} {\bibfnamefont {A.}~\bibnamefont {Hasenfratz}},
  \bibinfo {author} {\bibfnamefont {C.~E.}\ \bibnamefont {Detar}}, \bibinfo
  {author} {\bibfnamefont {J.}~\bibnamefont {Osborn}}, \bibinfo {author}
  {\bibfnamefont {S.~A.}\ \bibnamefont {Gottlieb}}, \bibinfo {author}
  {\bibfnamefont {E.}~\bibnamefont {Gregory}}, \bibinfo {author} {\bibfnamefont
  {D.}~\bibnamefont {Toussaint}}, \bibinfo {author} {\bibfnamefont
  {A.}~\bibnamefont {Hart}}, \bibinfo {author} {\bibfnamefont {U.~M.}\
  \bibnamefont {Heller}}, \bibinfo {author} {\bibfnamefont {J.}~\bibnamefont
  {Hetrick}}, \ and\ \bibinfo {author} {\bibfnamefont {R.~L.}\ \bibnamefont
  {Sugar}},\ }\href {\doibase 10.1103/PhysRevD.68.114501} {\bibfield  {journal}
  {\bibinfo  {journal} {Phys. Rev.}\ }\textbf {\bibinfo {volume} {D68}},\
  \bibinfo {pages} {114501} (\bibinfo {year} {2003})},\ \Eprint
  {http://arxiv.org/abs/hep-lat/0308019} {arXiv:hep-lat/0308019 [hep-lat]}
  \BibitemShut {NoStop}%
\bibitem [{\citenamefont {Del~Debbio}\ \emph {et~al.}(2005)\citenamefont
  {Del~Debbio}, \citenamefont {Giusti},\ and\ \citenamefont
  {Pica}}]{DelDebbio:2004ns}%
  \BibitemOpen
  \bibfield  {author} {\bibinfo {author} {\bibfnamefont {L.}~\bibnamefont
  {Del~Debbio}}, \bibinfo {author} {\bibfnamefont {L.}~\bibnamefont {Giusti}},
  \ and\ \bibinfo {author} {\bibfnamefont {C.}~\bibnamefont {Pica}},\ }\href
  {\doibase 10.1103/PhysRevLett.94.032003} {\bibfield  {journal} {\bibinfo
  {journal} {Phys. Rev. Lett.}\ }\textbf {\bibinfo {volume} {94}},\ \bibinfo
  {pages} {032003} (\bibinfo {year} {2005})},\ \Eprint
  {http://arxiv.org/abs/hep-th/0407052} {arXiv:hep-th/0407052 [hep-th]}
  \BibitemShut {NoStop}%
\bibitem [{\citenamefont {D{\"u}rr}\ \emph {et~al.}(2007)\citenamefont
  {D{\"u}rr}, \citenamefont {Fodor}, \citenamefont {Hoelbling},\ and\
  \citenamefont {Kurth}}]{Durr:2006ky}%
  \BibitemOpen
  \bibfield  {author} {\bibinfo {author} {\bibfnamefont {S.}~\bibnamefont
  {D{\"u}rr}}, \bibinfo {author} {\bibfnamefont {Z.}~\bibnamefont {Fodor}},
  \bibinfo {author} {\bibfnamefont {C.}~\bibnamefont {Hoelbling}}, \ and\
  \bibinfo {author} {\bibfnamefont {T.}~\bibnamefont {Kurth}},\ }\href
  {\doibase 10.1088/1126-6708/2007/04/055} {\bibfield  {journal} {\bibinfo
  {journal} {JHEP}\ }\textbf {\bibinfo {volume} {04}},\ \bibinfo {pages} {055}
  (\bibinfo {year} {2007})},\ \Eprint {http://arxiv.org/abs/hep-lat/0612021}
  {arXiv:hep-lat/0612021 [hep-lat]} \BibitemShut {NoStop}%
\bibitem [{\citenamefont {Bruno}\ \emph {et~al.}(2014)\citenamefont {Bruno},
  \citenamefont {Schaefer},\ and\ \citenamefont {Sommer}}]{Bruno:2014ova}%
  \BibitemOpen
  \bibfield  {author} {\bibinfo {author} {\bibfnamefont {M.}~\bibnamefont
  {Bruno}}, \bibinfo {author} {\bibfnamefont {S.}~\bibnamefont {Schaefer}}, \
  and\ \bibinfo {author} {\bibfnamefont {R.}~\bibnamefont {Sommer}} (\bibinfo
  {collaboration} {ALPHA}),\ }\href {\doibase 10.1007/JHEP08(2014)150}
  {\bibfield  {journal} {\bibinfo  {journal} {JHEP}\ }\textbf {\bibinfo
  {volume} {08}},\ \bibinfo {pages} {150} (\bibinfo {year} {2014})},\ \Eprint
  {http://arxiv.org/abs/1406.5363} {arXiv:1406.5363 [hep-lat]} \BibitemShut
  {NoStop}%
\bibitem [{\citenamefont {Aoki}\ \emph {et~al.}(2018)\citenamefont {Aoki},
  \citenamefont {Cossu}, \citenamefont {Fukaya}, \citenamefont {Hashimoto},\
  and\ \citenamefont {Kaneko}}]{Aoki:2017paw}%
  \BibitemOpen
  \bibfield  {author} {\bibinfo {author} {\bibfnamefont {S.}~\bibnamefont
  {Aoki}}, \bibinfo {author} {\bibfnamefont {G.}~\bibnamefont {Cossu}},
  \bibinfo {author} {\bibfnamefont {H.}~\bibnamefont {Fukaya}}, \bibinfo
  {author} {\bibfnamefont {S.}~\bibnamefont {Hashimoto}}, \ and\ \bibinfo
  {author} {\bibfnamefont {T.}~\bibnamefont {Kaneko}} (\bibinfo {collaboration}
  {JLQCD}),\ }\href {\doibase 10.1093/ptep/pty041} {\bibfield  {journal}
  {\bibinfo  {journal} {PTEP}\ }\textbf {\bibinfo {volume} {2018}},\ \bibinfo
  {pages} {043B07} (\bibinfo {year} {2018})},\ \Eprint
  {http://arxiv.org/abs/1705.10906} {arXiv:1705.10906 [hep-lat]} \BibitemShut
  {NoStop}%
\bibitem [{\citenamefont {Leutwyler}\ and\ \citenamefont
  {Smilga}(1992)}]{Leutwyler:1992yt}%
  \BibitemOpen
  \bibfield  {author} {\bibinfo {author} {\bibfnamefont {H.}~\bibnamefont
  {Leutwyler}}\ and\ \bibinfo {author} {\bibfnamefont {A.~V.}\ \bibnamefont
  {Smilga}},\ }\href {\doibase 10.1103/PhysRevD.46.5607} {\bibfield  {journal}
  {\bibinfo  {journal} {Phys. Rev.}\ }\textbf {\bibinfo {volume} {D46}},\
  \bibinfo {pages} {5607} (\bibinfo {year} {1992})}\BibitemShut {NoStop}%
\bibitem [{\citenamefont {Hasenfratz}\ \emph {et~al.}(1998)\citenamefont
  {Hasenfratz}, \citenamefont {Laliena},\ and\ \citenamefont
  {Niedermayer}}]{Hasenfratz:1998ri}%
  \BibitemOpen
  \bibfield  {author} {\bibinfo {author} {\bibfnamefont {P.}~\bibnamefont
  {Hasenfratz}}, \bibinfo {author} {\bibfnamefont {V.}~\bibnamefont {Laliena}},
  \ and\ \bibinfo {author} {\bibfnamefont {F.}~\bibnamefont {Niedermayer}},\
  }\href {\doibase 10.1016/S0370-2693(98)00315-3} {\bibfield  {journal}
  {\bibinfo  {journal} {Phys.\ Lett.\ B}\ }\textbf {\bibinfo {volume} {427}},\
  \bibinfo {pages} {125} (\bibinfo {year} {1998})},\ \Eprint
  {http://arxiv.org/abs/hep-lat/9801021} {arXiv:hep-lat/9801021} \BibitemShut
  {NoStop}%
\bibitem [{\citenamefont {L{\"u}scher}(2010)}]{Luscher:2009eq}%
  \BibitemOpen
  \bibfield  {author} {\bibinfo {author} {\bibfnamefont {M.}~\bibnamefont
  {L{\"u}scher}},\ }\href {\doibase 10.1007/s00220-009-0953-7} {\bibfield
  {journal} {\bibinfo  {journal} {Commun.Math.Phys.}\ }\textbf {\bibinfo
  {volume} {293}},\ \bibinfo {pages} {899} (\bibinfo {year} {2010})},\ \Eprint
  {http://arxiv.org/abs/0907.5491} {arXiv:0907.5491 [hep-lat]} \BibitemShut
  {NoStop}%
\bibitem [{\citenamefont {L{\"{u}}scher}(2010)}]{Luscher:2010iy}%
  \BibitemOpen
  \bibfield  {author} {\bibinfo {author} {\bibfnamefont {M.}~\bibnamefont
  {L{\"{u}}scher}},\ }\href {\doibase 10.1007/JHEP08(2010)071} {\bibfield
  {journal} {\bibinfo  {journal} {JHEP}\ }\textbf {\bibinfo {volume} {1008}},\
  \bibinfo {pages} {071} (\bibinfo {year} {2010})},\ \Eprint
  {http://arxiv.org/abs/1006.4518} {arXiv:1006.4518 [hep-lat]} \BibitemShut
  {NoStop}%
\bibitem [{\citenamefont {Narayanan}\ and\ \citenamefont
  {Neuberger}(2006)}]{Narayanan:2006rf}%
  \BibitemOpen
  \bibfield  {author} {\bibinfo {author} {\bibfnamefont {R.}~\bibnamefont
  {Narayanan}}\ and\ \bibinfo {author} {\bibfnamefont {H.}~\bibnamefont
  {Neuberger}},\ }\href {\doibase 10.1088/1126-6708/2006/03/064} {\bibfield
  {journal} {\bibinfo  {journal} {JHEP}\ }\textbf {\bibinfo {volume} {0603}},\
  \bibinfo {pages} {064} (\bibinfo {year} {2006})},\ \Eprint
  {http://arxiv.org/abs/hep-th/0601210} {arXiv:hep-th/0601210 [hep-th]}
  \BibitemShut {NoStop}%
\bibitem [{\citenamefont {Alexandrou}\ \emph {et~al.}(2015)\citenamefont
  {Alexandrou}, \citenamefont {Athenodorou},\ and\ \citenamefont
  {Jansen}}]{Alexandrou:2015yba}%
  \BibitemOpen
  \bibfield  {author} {\bibinfo {author} {\bibfnamefont {C.}~\bibnamefont
  {Alexandrou}}, \bibinfo {author} {\bibfnamefont {A.}~\bibnamefont
  {Athenodorou}}, \ and\ \bibinfo {author} {\bibfnamefont {K.}~\bibnamefont
  {Jansen}},\ }\href {\doibase 10.1103/PhysRevD.92.125014} {\bibfield
  {journal} {\bibinfo  {journal} {Phys. Rev.}\ }\textbf {\bibinfo {volume}
  {D92}},\ \bibinfo {pages} {125014} (\bibinfo {year} {2015})},\ \Eprint
  {http://arxiv.org/abs/1509.04259} {arXiv:1509.04259 [hep-lat]} \BibitemShut
  {NoStop}%
\bibitem [{\citenamefont {Alexandrou}\ \emph {et~al.}(2017)\citenamefont
  {Alexandrou}, \citenamefont {Athenodorou}, \citenamefont {Cichy},
  \citenamefont {Dromard}, \citenamefont {Garcia-Ramos}, \citenamefont
  {Jansen}, \citenamefont {Wenger},\ and\ \citenamefont
  {Zimmermann}}]{Alexandrou:2017hqw}%
  \BibitemOpen
  \bibfield  {author} {\bibinfo {author} {\bibfnamefont {C.}~\bibnamefont
  {Alexandrou}}, \bibinfo {author} {\bibfnamefont {A.}~\bibnamefont
  {Athenodorou}}, \bibinfo {author} {\bibfnamefont {K.}~\bibnamefont {Cichy}},
  \bibinfo {author} {\bibfnamefont {A.}~\bibnamefont {Dromard}}, \bibinfo
  {author} {\bibfnamefont {E.}~\bibnamefont {Garcia-Ramos}}, \bibinfo {author}
  {\bibfnamefont {K.}~\bibnamefont {Jansen}}, \bibinfo {author} {\bibfnamefont
  {U.}~\bibnamefont {Wenger}}, \ and\ \bibinfo {author} {\bibfnamefont
  {F.}~\bibnamefont {Zimmermann}},\ }\href@noop {} {\  (\bibinfo {year}
  {2017})},\ \Eprint {http://arxiv.org/abs/1708.00696} {arXiv:1708.00696
  [hep-lat]} \BibitemShut {NoStop}%
\bibitem [{\citenamefont {Fodor}\ \emph
  {et~al.}(2012{\natexlab{a}})\citenamefont {Fodor}, \citenamefont {Holland},
  \citenamefont {Kuti}, \citenamefont {Nogradi},\ and\ \citenamefont
  {Wong}}]{Fodor:2012td}%
  \BibitemOpen
  \bibfield  {author} {\bibinfo {author} {\bibfnamefont {Z.}~\bibnamefont
  {Fodor}}, \bibinfo {author} {\bibfnamefont {K.}~\bibnamefont {Holland}},
  \bibinfo {author} {\bibfnamefont {J.}~\bibnamefont {Kuti}}, \bibinfo {author}
  {\bibfnamefont {D.}~\bibnamefont {Nogradi}}, \ and\ \bibinfo {author}
  {\bibfnamefont {C.~H.}\ \bibnamefont {Wong}},\ }\href {\doibase
  10.1007/JHEP11(2012)007} {\bibfield  {journal} {\bibinfo  {journal} {JHEP}\
  }\textbf {\bibinfo {volume} {1211}},\ \bibinfo {pages} {007} (\bibinfo {year}
  {2012}{\natexlab{a}})},\ \Eprint {http://arxiv.org/abs/1208.1051}
  {arXiv:1208.1051 [hep-lat]} \BibitemShut {NoStop}%
\bibitem [{\citenamefont {Fodor}\ \emph {et~al.}(2014)\citenamefont {Fodor},
  \citenamefont {Holland}, \citenamefont {Kuti}, \citenamefont {Mondal},
  \citenamefont {Nogradi},\ and\ \citenamefont {Wong}}]{Fodor:2014cpa}%
  \BibitemOpen
  \bibfield  {author} {\bibinfo {author} {\bibfnamefont {Z.}~\bibnamefont
  {Fodor}}, \bibinfo {author} {\bibfnamefont {K.}~\bibnamefont {Holland}},
  \bibinfo {author} {\bibfnamefont {J.}~\bibnamefont {Kuti}}, \bibinfo {author}
  {\bibfnamefont {S.}~\bibnamefont {Mondal}}, \bibinfo {author} {\bibfnamefont
  {D.}~\bibnamefont {Nogradi}}, \ and\ \bibinfo {author} {\bibfnamefont
  {C.~H.}\ \bibnamefont {Wong}},\ }\href {\doibase 10.1007/JHEP09(2014)018}
  {\bibfield  {journal} {\bibinfo  {journal} {JHEP}\ }\textbf {\bibinfo
  {volume} {09}},\ \bibinfo {pages} {018} (\bibinfo {year} {2014})},\ \Eprint
  {http://arxiv.org/abs/1406.0827} {arXiv:1406.0827 [hep-lat]} \BibitemShut
  {NoStop}%
\bibitem [{\citenamefont {Hasenfratz}\ \emph {et~al.}(2015)\citenamefont
  {Hasenfratz}, \citenamefont {Schaich},\ and\ \citenamefont
  {Veernala}}]{Hasenfratz:2014rna}%
  \BibitemOpen
  \bibfield  {author} {\bibinfo {author} {\bibfnamefont {A.}~\bibnamefont
  {Hasenfratz}}, \bibinfo {author} {\bibfnamefont {D.}~\bibnamefont {Schaich}},
  \ and\ \bibinfo {author} {\bibfnamefont {A.}~\bibnamefont {Veernala}},\
  }\href {\doibase 10.1007/JHEP06(2015)143} {\bibfield  {journal} {\bibinfo
  {journal} {JHEP}\ }\textbf {\bibinfo {volume} {06}},\ \bibinfo {pages} {143}
  (\bibinfo {year} {2015})},\ \Eprint {http://arxiv.org/abs/1410.5886}
  {arXiv:1410.5886 [hep-lat]} \BibitemShut {NoStop}%
\bibitem [{\citenamefont {Dalla~Brida}\ \emph {et~al.}(2017)\citenamefont
  {Dalla~Brida}, \citenamefont {Fritzsch}, \citenamefont {Korzec},
  \citenamefont {Ramos}, \citenamefont {Sint},\ and\ \citenamefont
  {Sommer}}]{DallaBrida:2016kgh}%
  \BibitemOpen
  \bibfield  {author} {\bibinfo {author} {\bibfnamefont {M.}~\bibnamefont
  {Dalla~Brida}}, \bibinfo {author} {\bibfnamefont {P.}~\bibnamefont
  {Fritzsch}}, \bibinfo {author} {\bibfnamefont {T.}~\bibnamefont {Korzec}},
  \bibinfo {author} {\bibfnamefont {A.}~\bibnamefont {Ramos}}, \bibinfo
  {author} {\bibfnamefont {S.}~\bibnamefont {Sint}}, \ and\ \bibinfo {author}
  {\bibfnamefont {R.}~\bibnamefont {Sommer}} (\bibinfo {collaboration}
  {ALPHA}),\ }\href {\doibase 10.1103/PhysRevD.95.014507} {\bibfield  {journal}
  {\bibinfo  {journal} {Phys. Rev.}\ }\textbf {\bibinfo {volume} {D95}},\
  \bibinfo {pages} {014507} (\bibinfo {year} {2017})},\ \Eprint
  {http://arxiv.org/abs/1607.06423} {arXiv:1607.06423 [hep-lat]} \BibitemShut
  {NoStop}%
\bibitem [{\citenamefont {Ramos}\ and\ \citenamefont
  {Sint}(2016)}]{Ramos:2015baa}%
  \BibitemOpen
  \bibfield  {author} {\bibinfo {author} {\bibfnamefont {A.}~\bibnamefont
  {Ramos}}\ and\ \bibinfo {author} {\bibfnamefont {S.}~\bibnamefont {Sint}},\
  }\href {\doibase 10.1140/epjc/s10052-015-3831-9} {\bibfield  {journal}
  {\bibinfo  {journal} {Eur. Phys. J.}\ }\textbf {\bibinfo {volume} {C76}},\
  \bibinfo {pages} {15} (\bibinfo {year} {2016})},\ \Eprint
  {http://arxiv.org/abs/1508.05552} {arXiv:1508.05552 [hep-lat]} \BibitemShut
  {NoStop}%
\bibitem [{\citenamefont {Hasenfratz}\ and\ \citenamefont
  {Schaich}(2018)}]{Hasenfratz:2016dou}%
  \BibitemOpen
  \bibfield  {author} {\bibinfo {author} {\bibfnamefont {A.}~\bibnamefont
  {Hasenfratz}}\ and\ \bibinfo {author} {\bibfnamefont {D.}~\bibnamefont
  {Schaich}},\ }\href {\doibase 10.1007/JHEP02(2018)132} {\bibfield  {journal}
  {\bibinfo  {journal} {JHEP}\ }\textbf {\bibinfo {volume} {02}},\ \bibinfo
  {pages} {132} (\bibinfo {year} {2018})},\ \Eprint
  {http://arxiv.org/abs/1610.10004} {arXiv:1610.10004 [hep-lat]} \BibitemShut
  {NoStop}%
\bibitem [{\citenamefont {Hasenfratz}\ \emph
  {et~al.}(2019{\natexlab{a}})\citenamefont {Hasenfratz}, \citenamefont
  {Rebbi},\ and\ \citenamefont {Witzel}}]{Hasenfratz:2017qyr}%
  \BibitemOpen
  \bibfield  {author} {\bibinfo {author} {\bibfnamefont {A.}~\bibnamefont
  {Hasenfratz}}, \bibinfo {author} {\bibfnamefont {C.}~\bibnamefont {Rebbi}}, \
  and\ \bibinfo {author} {\bibfnamefont {O.}~\bibnamefont {Witzel}},\ }\href
  {\doibase 10.1016/j.physletb.2019.134937} {\bibfield  {journal} {\bibinfo
  {journal} {Phys. Lett.}\ }\textbf {\bibinfo {volume} {B798}},\ \bibinfo
  {pages} {134937} (\bibinfo {year} {2019}{\natexlab{a}})},\ \Eprint
  {http://arxiv.org/abs/1710.11578} {arXiv:1710.11578 [hep-lat]} \BibitemShut
  {NoStop}%
\bibitem [{\citenamefont {Hasenfratz}\ \emph
  {et~al.}(2019{\natexlab{b}})\citenamefont {Hasenfratz}, \citenamefont
  {Rebbi},\ and\ \citenamefont {Witzel}}]{Hasenfratz:2019dpr}%
  \BibitemOpen
  \bibfield  {author} {\bibinfo {author} {\bibfnamefont {A.}~\bibnamefont
  {Hasenfratz}}, \bibinfo {author} {\bibfnamefont {C.}~\bibnamefont {Rebbi}}, \
  and\ \bibinfo {author} {\bibfnamefont {O.}~\bibnamefont {Witzel}},\ }\href
  {\doibase 10.1103/PhysRevD.100.114508} {\bibfield  {journal} {\bibinfo
  {journal} {Phys. Rev.}\ }\textbf {\bibinfo {volume} {D100}},\ \bibinfo
  {pages} {114508} (\bibinfo {year} {2019}{\natexlab{b}})},\ \Eprint
  {http://arxiv.org/abs/1909.05842} {arXiv:1909.05842 [hep-lat]} \BibitemShut
  {NoStop}%
\bibitem [{\citenamefont {Fodor}\ \emph
  {et~al.}(2018{\natexlab{a}})\citenamefont {Fodor}, \citenamefont {Holland},
  \citenamefont {Kuti}, \citenamefont {Nogradi},\ and\ \citenamefont
  {Wong}}]{Fodor:2017gtj}%
  \BibitemOpen
  \bibfield  {author} {\bibinfo {author} {\bibfnamefont {Z.}~\bibnamefont
  {Fodor}}, \bibinfo {author} {\bibfnamefont {K.}~\bibnamefont {Holland}},
  \bibinfo {author} {\bibfnamefont {J.}~\bibnamefont {Kuti}}, \bibinfo {author}
  {\bibfnamefont {D.}~\bibnamefont {Nogradi}}, \ and\ \bibinfo {author}
  {\bibfnamefont {C.~H.}\ \bibnamefont {Wong}},\ }\href {\doibase
  10.1016/j.physletb.2018.02.008} {\bibfield  {journal} {\bibinfo  {journal}
  {Phys. Lett.}\ }\textbf {\bibinfo {volume} {B779}},\ \bibinfo {pages} {230}
  (\bibinfo {year} {2018}{\natexlab{a}})},\ \Eprint
  {http://arxiv.org/abs/1710.09262} {arXiv:1710.09262 [hep-lat]} \BibitemShut
  {NoStop}%
\bibitem [{\citenamefont {Fodor}\ \emph {et~al.}(2019)\citenamefont {Fodor},
  \citenamefont {Holland}, \citenamefont {Kuti}, \citenamefont {Nogradi},\ and\
  \citenamefont {Wong}}]{Fodor:2019ypi}%
  \BibitemOpen
  \bibfield  {author} {\bibinfo {author} {\bibfnamefont {Z.}~\bibnamefont
  {Fodor}}, \bibinfo {author} {\bibfnamefont {K.}~\bibnamefont {Holland}},
  \bibinfo {author} {\bibfnamefont {J.}~\bibnamefont {Kuti}}, \bibinfo {author}
  {\bibfnamefont {D.}~\bibnamefont {Nogradi}}, \ and\ \bibinfo {author}
  {\bibfnamefont {C.~H.}\ \bibnamefont {Wong}},\ }\href@noop {} {\  (\bibinfo
  {year} {2019})},\ \Eprint {http://arxiv.org/abs/1912.07653} {arXiv:1912.07653
  [hep-lat]} \BibitemShut {NoStop}%
\bibitem [{\citenamefont {Borsanyi}\ \emph {et~al.}(2012)\citenamefont
  {Borsanyi}, \citenamefont {D{\"u}rr}, \citenamefont {Fodor}, \citenamefont
  {Hoelbling}, \citenamefont {Katz}, \citenamefont {Krieg}, \citenamefont
  {Kurth}, \citenamefont {Lellouch}, \citenamefont {Lippert}, \citenamefont
  {McNeile},\ and\ \citenamefont {Szabo}}]{Borsanyi:2012zs}%
  \BibitemOpen
  \bibfield  {author} {\bibinfo {author} {\bibfnamefont {S.}~\bibnamefont
  {Borsanyi}}, \bibinfo {author} {\bibfnamefont {S.}~\bibnamefont {D{\"u}rr}},
  \bibinfo {author} {\bibfnamefont {Z.}~\bibnamefont {Fodor}}, \bibinfo
  {author} {\bibfnamefont {C.}~\bibnamefont {Hoelbling}}, \bibinfo {author}
  {\bibfnamefont {S.~D.}\ \bibnamefont {Katz}}, \bibinfo {author}
  {\bibfnamefont {S.}~\bibnamefont {Krieg}}, \bibinfo {author} {\bibfnamefont
  {T.}~\bibnamefont {Kurth}}, \bibinfo {author} {\bibfnamefont
  {L.}~\bibnamefont {Lellouch}}, \bibinfo {author} {\bibfnamefont
  {T.}~\bibnamefont {Lippert}}, \bibinfo {author} {\bibfnamefont
  {C.}~\bibnamefont {McNeile}}, \ and\ \bibinfo {author} {\bibfnamefont
  {K.~K.}\ \bibnamefont {Szabo}},\ }\href {\doibase 10.1007/JHEP09(2012)010}
  {\bibfield  {journal} {\bibinfo  {journal} {JHEP}\ }\textbf {\bibinfo
  {volume} {09}},\ \bibinfo {pages} {010} (\bibinfo {year} {2012})},\ \Eprint
  {http://arxiv.org/abs/1203.4469} {arXiv:1203.4469 [hep-lat]} \BibitemShut
  {NoStop}%
\bibitem [{\citenamefont {Sommer}(2014)}]{Sommer:2014mea}%
  \BibitemOpen
  \bibfield  {author} {\bibinfo {author} {\bibfnamefont {R.}~\bibnamefont
  {Sommer}},\ }\href {\doibase 10.22323/1.187.0015} {\bibfield  {journal}
  {\bibinfo  {journal} {PoS}\ }\textbf {\bibinfo {volume} {LATTICE2013}},\
  \bibinfo {pages} {015} (\bibinfo {year} {2014})},\ \Eprint
  {http://arxiv.org/abs/1401.3270} {arXiv:1401.3270 [hep-lat]} \BibitemShut
  {NoStop}%
\bibitem [{\citenamefont {Witzel}(2019)}]{Witzel:2019jbe}%
  \BibitemOpen
  \bibfield  {author} {\bibinfo {author} {\bibfnamefont {O.}~\bibnamefont
  {Witzel}},\ }\href {\doibase 10.22323/1.334.0006} {\bibfield  {journal}
  {\bibinfo  {journal} {PoS}\ }\textbf {\bibinfo {volume} {LATTICE2018}},\
  \bibinfo {pages} {006} (\bibinfo {year} {2019})},\ \Eprint
  {http://arxiv.org/abs/1901.08216} {arXiv:1901.08216 [hep-lat]} \BibitemShut
  {NoStop}%
\bibitem [{\citenamefont {Brower}\ \emph {et~al.}(2019)\citenamefont {Brower},
  \citenamefont {Hasenfratz}, \citenamefont {Neil}, \citenamefont {Catterall},
  \citenamefont {Fleming}, \citenamefont {Giedt}, \citenamefont {Rinaldi},
  \citenamefont {Schaich}, \citenamefont {Weinberg},\ and\ \citenamefont
  {Witzel}}]{Brower:2019oor}%
  \BibitemOpen
  \bibfield  {author} {\bibinfo {author} {\bibfnamefont {R.~C.}\ \bibnamefont
  {Brower}}, \bibinfo {author} {\bibfnamefont {A.}~\bibnamefont {Hasenfratz}},
  \bibinfo {author} {\bibfnamefont {E.~T.}\ \bibnamefont {Neil}}, \bibinfo
  {author} {\bibfnamefont {S.}~\bibnamefont {Catterall}}, \bibinfo {author}
  {\bibfnamefont {G.}~\bibnamefont {Fleming}}, \bibinfo {author} {\bibfnamefont
  {J.}~\bibnamefont {Giedt}}, \bibinfo {author} {\bibfnamefont
  {E.}~\bibnamefont {Rinaldi}}, \bibinfo {author} {\bibfnamefont
  {D.}~\bibnamefont {Schaich}}, \bibinfo {author} {\bibfnamefont
  {E.}~\bibnamefont {Weinberg}}, \ and\ \bibinfo {author} {\bibfnamefont
  {O.}~\bibnamefont {Witzel}} (\bibinfo {collaboration} {USQCD}),\ }\href
  {\doibase 10.1140/epja/i2019-12901-5} {\bibfield  {journal} {\bibinfo
  {journal} {Eur. Phys. J.}\ }\textbf {\bibinfo {volume} {A55}},\ \bibinfo
  {pages} {198} (\bibinfo {year} {2019})},\ \Eprint
  {http://arxiv.org/abs/1904.09964} {arXiv:1904.09964 [hep-lat]} \BibitemShut
  {NoStop}%
\bibitem [{\citenamefont {Fodor}\ \emph
  {et~al.}(2012{\natexlab{b}})\citenamefont {Fodor}, \citenamefont {Holland},
  \citenamefont {Kuti}, \citenamefont {Nogradi}, \citenamefont {Schroeder},\
  and\ \citenamefont {Wong}}]{Fodor:2012ty}%
  \BibitemOpen
  \bibfield  {author} {\bibinfo {author} {\bibfnamefont {Z.}~\bibnamefont
  {Fodor}}, \bibinfo {author} {\bibfnamefont {K.}~\bibnamefont {Holland}},
  \bibinfo {author} {\bibfnamefont {J.}~\bibnamefont {Kuti}}, \bibinfo {author}
  {\bibfnamefont {D.}~\bibnamefont {Nogradi}}, \bibinfo {author} {\bibfnamefont
  {C.}~\bibnamefont {Schroeder}}, \ and\ \bibinfo {author} {\bibfnamefont
  {C.~H.}\ \bibnamefont {Wong}},\ }\href {\doibase
  10.1016/j.physletb.2012.10.079} {\bibfield  {journal} {\bibinfo  {journal}
  {Phys. Lett.}\ }\textbf {\bibinfo {volume} {B718}},\ \bibinfo {pages} {657}
  (\bibinfo {year} {2012}{\natexlab{b}})},\ \Eprint
  {http://arxiv.org/abs/1209.0391} {arXiv:1209.0391 [hep-lat]} \BibitemShut
  {NoStop}%
\bibitem [{\citenamefont {Fodor}\ \emph {et~al.}(2015)\citenamefont {Fodor},
  \citenamefont {Holland}, \citenamefont {Kuti}, \citenamefont {Mondal},
  \citenamefont {Nogradi},\ and\ \citenamefont {Wong}}]{Fodor:2015zna}%
  \BibitemOpen
  \bibfield  {author} {\bibinfo {author} {\bibfnamefont {Z.}~\bibnamefont
  {Fodor}}, \bibinfo {author} {\bibfnamefont {K.}~\bibnamefont {Holland}},
  \bibinfo {author} {\bibfnamefont {J.}~\bibnamefont {Kuti}}, \bibinfo {author}
  {\bibfnamefont {S.}~\bibnamefont {Mondal}}, \bibinfo {author} {\bibfnamefont
  {D.}~\bibnamefont {Nogradi}}, \ and\ \bibinfo {author} {\bibfnamefont
  {C.~H.}\ \bibnamefont {Wong}},\ }\href {\doibase 10.1007/JHEP09(2015)039}
  {\bibfield  {journal} {\bibinfo  {journal} {JHEP}\ }\textbf {\bibinfo
  {volume} {09}},\ \bibinfo {pages} {039} (\bibinfo {year} {2015})},\ \Eprint
  {http://arxiv.org/abs/1506.06599} {arXiv:1506.06599 [hep-lat]} \BibitemShut
  {NoStop}%
\bibitem [{\citenamefont {Fodor}\ \emph {et~al.}(2016)\citenamefont {Fodor},
  \citenamefont {Holland}, \citenamefont {Kuti}, \citenamefont {Mondal},
  \citenamefont {Nogradi},\ and\ \citenamefont {Wong}}]{Fodor:2016wal}%
  \BibitemOpen
  \bibfield  {author} {\bibinfo {author} {\bibfnamefont {Z.}~\bibnamefont
  {Fodor}}, \bibinfo {author} {\bibfnamefont {K.}~\bibnamefont {Holland}},
  \bibinfo {author} {\bibfnamefont {J.}~\bibnamefont {Kuti}}, \bibinfo {author}
  {\bibfnamefont {S.}~\bibnamefont {Mondal}}, \bibinfo {author} {\bibfnamefont
  {D.}~\bibnamefont {Nogradi}}, \ and\ \bibinfo {author} {\bibfnamefont
  {C.~H.}\ \bibnamefont {Wong}},\ }\href {\doibase 10.1103/PhysRevD.94.014503}
  {\bibfield  {journal} {\bibinfo  {journal} {Phys. Rev.}\ }\textbf {\bibinfo
  {volume} {D94}},\ \bibinfo {pages} {014503} (\bibinfo {year} {2016})},\
  \Eprint {http://arxiv.org/abs/1601.03302} {arXiv:1601.03302 [hep-lat]}
  \BibitemShut {NoStop}%
\bibitem [{\citenamefont {Appelquist}\ \emph {et~al.}(2014)\citenamefont
  {Appelquist}, \citenamefont {Brower}, \citenamefont {Fleming}, \citenamefont
  {Kiskis}, \citenamefont {Lin}, \citenamefont {Neil}, \citenamefont {Osborn},
  \citenamefont {Rebbi}, \citenamefont {Rinaldi}, \citenamefont {Schaich},
  \citenamefont {Schroeder}, \citenamefont {Syritsyn}, \citenamefont {Voronov},
  \citenamefont {Vranas}, \citenamefont {Weinberg},\ and\ \citenamefont
  {Witzel}}]{Appelquist:2014zsa}%
  \BibitemOpen
  \bibfield  {author} {\bibinfo {author} {\bibfnamefont {T.}~\bibnamefont
  {Appelquist}}, \bibinfo {author} {\bibfnamefont {R.}~\bibnamefont {Brower}},
  \bibinfo {author} {\bibfnamefont {G.}~\bibnamefont {Fleming}}, \bibinfo
  {author} {\bibfnamefont {J.}~\bibnamefont {Kiskis}}, \bibinfo {author}
  {\bibfnamefont {M.}~\bibnamefont {Lin}}, \bibinfo {author} {\bibfnamefont
  {E.}~\bibnamefont {Neil}}, \bibinfo {author} {\bibfnamefont {J.}~\bibnamefont
  {Osborn}}, \bibinfo {author} {\bibfnamefont {C.}~\bibnamefont {Rebbi}},
  \bibinfo {author} {\bibfnamefont {E.}~\bibnamefont {Rinaldi}}, \bibinfo
  {author} {\bibfnamefont {D.}~\bibnamefont {Schaich}}, \bibinfo {author}
  {\bibfnamefont {C.}~\bibnamefont {Schroeder}}, \bibinfo {author}
  {\bibfnamefont {S.}~\bibnamefont {Syritsyn}}, \bibinfo {author}
  {\bibfnamefont {G.}~\bibnamefont {Voronov}}, \bibinfo {author} {\bibfnamefont
  {P.}~\bibnamefont {Vranas}}, \bibinfo {author} {\bibfnamefont
  {E.}~\bibnamefont {Weinberg}}, \ and\ \bibinfo {author} {\bibfnamefont
  {O.}~\bibnamefont {Witzel}} (\bibinfo {collaboration} {Lattice Strong
  Dynamics}),\ }\href {\doibase 10.1103/PhysRevD.90.114502} {\bibfield
  {journal} {\bibinfo  {journal} {Phys. Rev.}\ }\textbf {\bibinfo {volume}
  {D90}},\ \bibinfo {pages} {114502} (\bibinfo {year} {2014})},\ \Eprint
  {http://arxiv.org/abs/1405.4752} {arXiv:1405.4752 [hep-lat]} \BibitemShut
  {NoStop}%
\bibitem [{\citenamefont {Appelquist}\ \emph {et~al.}(2019)\citenamefont
  {Appelquist}, \citenamefont {Brower}, \citenamefont {Fleming}, \citenamefont
  {Gasbarro}, \citenamefont {Hasenfratz}, \citenamefont {Jin}, \citenamefont
  {Neil}, \citenamefont {Osborn}, \citenamefont {Rebbi}, \citenamefont
  {Rinaldi}, \citenamefont {Schaich}, \citenamefont {Vranas}, \citenamefont
  {Weinberg},\ and\ \citenamefont {Witzel}}]{Appelquist:2018yqe}%
  \BibitemOpen
  \bibfield  {author} {\bibinfo {author} {\bibfnamefont {T.}~\bibnamefont
  {Appelquist}}, \bibinfo {author} {\bibfnamefont {R.}~\bibnamefont {Brower}},
  \bibinfo {author} {\bibfnamefont {G.}~\bibnamefont {Fleming}}, \bibinfo
  {author} {\bibfnamefont {A.}~\bibnamefont {Gasbarro}}, \bibinfo {author}
  {\bibfnamefont {A.}~\bibnamefont {Hasenfratz}}, \bibinfo {author}
  {\bibfnamefont {X.-Y.}\ \bibnamefont {Jin}}, \bibinfo {author} {\bibfnamefont
  {E.}~\bibnamefont {Neil}}, \bibinfo {author} {\bibfnamefont {J.}~\bibnamefont
  {Osborn}}, \bibinfo {author} {\bibfnamefont {C.}~\bibnamefont {Rebbi}},
  \bibinfo {author} {\bibfnamefont {E.}~\bibnamefont {Rinaldi}}, \bibinfo
  {author} {\bibfnamefont {D.}~\bibnamefont {Schaich}}, \bibinfo {author}
  {\bibfnamefont {P.}~\bibnamefont {Vranas}}, \bibinfo {author} {\bibfnamefont
  {E.}~\bibnamefont {Weinberg}}, \ and\ \bibinfo {author} {\bibfnamefont
  {O.}~\bibnamefont {Witzel}} (\bibinfo {collaboration} {Lattice Strong
  Dynamics}),\ }\href {\doibase 10.1103/PhysRevD.99.014509} {\bibfield
  {journal} {\bibinfo  {journal} {Phys. Rev.}\ }\textbf {\bibinfo {volume}
  {D99}},\ \bibinfo {pages} {014509} (\bibinfo {year} {2019})},\ \Eprint
  {http://arxiv.org/abs/1807.08411} {arXiv:1807.08411 [hep-lat]} \BibitemShut
  {NoStop}%
\bibitem [{\citenamefont {Aoki}\ \emph {et~al.}(2017)\citenamefont {Aoki},
  \citenamefont {Aoyama}, \citenamefont {Bennett}, \citenamefont {Kurachi},
  \citenamefont {Maskawa}, \citenamefont {Miura}, \citenamefont {Nagai},
  \citenamefont {Ohki}, \citenamefont {Rinaldi}, \citenamefont {Shibata},
  \citenamefont {Yamawaki},\ and\ \citenamefont {Yamazaki}}]{Aoki:2016wnc}%
  \BibitemOpen
  \bibfield  {author} {\bibinfo {author} {\bibfnamefont {Y.}~\bibnamefont
  {Aoki}}, \bibinfo {author} {\bibfnamefont {T.}~\bibnamefont {Aoyama}},
  \bibinfo {author} {\bibfnamefont {E.}~\bibnamefont {Bennett}}, \bibinfo
  {author} {\bibfnamefont {M.}~\bibnamefont {Kurachi}}, \bibinfo {author}
  {\bibfnamefont {T.}~\bibnamefont {Maskawa}}, \bibinfo {author} {\bibfnamefont
  {K.}~\bibnamefont {Miura}}, \bibinfo {author} {\bibfnamefont {K.-i.}\
  \bibnamefont {Nagai}}, \bibinfo {author} {\bibfnamefont {H.}~\bibnamefont
  {Ohki}}, \bibinfo {author} {\bibfnamefont {E.}~\bibnamefont {Rinaldi}},
  \bibinfo {author} {\bibfnamefont {A.}~\bibnamefont {Shibata}}, \bibinfo
  {author} {\bibfnamefont {K.}~\bibnamefont {Yamawaki}}, \ and\ \bibinfo
  {author} {\bibfnamefont {T.}~\bibnamefont {Yamazaki}} (\bibinfo
  {collaboration} {LatKMI}),\ }\href {\doibase 10.1103/PhysRevD.96.014508}
  {\bibfield  {journal} {\bibinfo  {journal} {Phys. Rev.}\ }\textbf {\bibinfo
  {volume} {D96}},\ \bibinfo {pages} {014508} (\bibinfo {year} {2017})},\
  \Eprint {http://arxiv.org/abs/1610.07011} {arXiv:1610.07011 [hep-lat]}
  \BibitemShut {NoStop}%
\bibitem [{\citenamefont {Aoki}\ \emph {et~al.}(2013)\citenamefont {Aoki},
  \citenamefont {Aoyama}, \citenamefont {Kurachi}, \citenamefont {Maskawa},
  \citenamefont {Nagai}, \citenamefont {Ohki}, \citenamefont {Shibata},
  \citenamefont {Yamawaki},\ and\ \citenamefont {Yamazaki}}]{Aoki:2013xza}%
  \BibitemOpen
  \bibfield  {author} {\bibinfo {author} {\bibfnamefont {Y.}~\bibnamefont
  {Aoki}}, \bibinfo {author} {\bibfnamefont {T.}~\bibnamefont {Aoyama}},
  \bibinfo {author} {\bibfnamefont {M.}~\bibnamefont {Kurachi}}, \bibinfo
  {author} {\bibfnamefont {T.}~\bibnamefont {Maskawa}}, \bibinfo {author}
  {\bibfnamefont {K.-i.}\ \bibnamefont {Nagai}}, \bibinfo {author}
  {\bibfnamefont {H.}~\bibnamefont {Ohki}}, \bibinfo {author} {\bibfnamefont
  {A.}~\bibnamefont {Shibata}}, \bibinfo {author} {\bibfnamefont
  {K.}~\bibnamefont {Yamawaki}}, \ and\ \bibinfo {author} {\bibfnamefont
  {T.}~\bibnamefont {Yamazaki}} (\bibinfo {collaboration} {LatKMI}),\ }\href
  {\doibase 10.1103/PhysRevD.87.094511} {\bibfield  {journal} {\bibinfo
  {journal} {Phys. Rev.}\ }\textbf {\bibinfo {volume} {D87}},\ \bibinfo {pages}
  {094511} (\bibinfo {year} {2013})},\ \Eprint {http://arxiv.org/abs/1302.6859}
  {arXiv:1302.6859 [hep-lat]} \BibitemShut {NoStop}%
\bibitem [{\citenamefont {Brower}\ \emph {et~al.}(2014)\citenamefont {Brower},
  \citenamefont {Hasenfratz}, \citenamefont {Rebbi}, \citenamefont {Weinberg},\
  and\ \citenamefont {Witzel}}]{Brower:2014ita}%
  \BibitemOpen
  \bibfield  {author} {\bibinfo {author} {\bibfnamefont {R.~C.}\ \bibnamefont
  {Brower}}, \bibinfo {author} {\bibfnamefont {A.}~\bibnamefont {Hasenfratz}},
  \bibinfo {author} {\bibfnamefont {C.}~\bibnamefont {Rebbi}}, \bibinfo
  {author} {\bibfnamefont {E.}~\bibnamefont {Weinberg}}, \ and\ \bibinfo
  {author} {\bibfnamefont {O.}~\bibnamefont {Witzel}},\ }\href@noop {}
  {\bibfield  {journal} {\bibinfo  {journal} {PoS}\ }\textbf {\bibinfo {volume}
  {LATTICE2014}},\ \bibinfo {pages} {254} (\bibinfo {year} {2014})},\ \Eprint
  {http://arxiv.org/abs/1411.3243} {arXiv:1411.3243 [hep-lat]} \BibitemShut
  {NoStop}%
\bibitem [{\citenamefont {Brower}\ \emph {et~al.}(2016)\citenamefont {Brower},
  \citenamefont {Hasenfratz}, \citenamefont {Rebbi}, \citenamefont {Weinberg},\
  and\ \citenamefont {Witzel}}]{Brower:2015owo}%
  \BibitemOpen
  \bibfield  {author} {\bibinfo {author} {\bibfnamefont {R.~C.}\ \bibnamefont
  {Brower}}, \bibinfo {author} {\bibfnamefont {A.}~\bibnamefont {Hasenfratz}},
  \bibinfo {author} {\bibfnamefont {C.}~\bibnamefont {Rebbi}}, \bibinfo
  {author} {\bibfnamefont {E.}~\bibnamefont {Weinberg}}, \ and\ \bibinfo
  {author} {\bibfnamefont {O.}~\bibnamefont {Witzel}},\ }\href {\doibase
  10.1103/PhysRevD.93.075028} {\bibfield  {journal} {\bibinfo  {journal} {Phys.
  Rev.}\ }\textbf {\bibinfo {volume} {D93}},\ \bibinfo {pages} {075028}
  (\bibinfo {year} {2016})},\ \Eprint {http://arxiv.org/abs/1512.02576}
  {arXiv:1512.02576 [hep-ph]} \BibitemShut {NoStop}%
\bibitem [{\citenamefont {Hasenfratz}\ \emph {et~al.}(2017)\citenamefont
  {Hasenfratz}, \citenamefont {Rebbi},\ and\ \citenamefont
  {Witzel}}]{Hasenfratz:2016gut}%
  \BibitemOpen
  \bibfield  {author} {\bibinfo {author} {\bibfnamefont {A.}~\bibnamefont
  {Hasenfratz}}, \bibinfo {author} {\bibfnamefont {C.}~\bibnamefont {Rebbi}}, \
  and\ \bibinfo {author} {\bibfnamefont {O.}~\bibnamefont {Witzel}},\ }\href
  {\doibase 10.1016/j.physletb.2017.07.058} {\bibfield  {journal} {\bibinfo
  {journal} {Phys. Lett.}\ }\textbf {\bibinfo {volume} {B773}},\ \bibinfo
  {pages} {86} (\bibinfo {year} {2017})},\ \Eprint
  {http://arxiv.org/abs/1609.01401} {arXiv:1609.01401 [hep-ph]} \BibitemShut
  {NoStop}%
\bibitem [{\citenamefont {Witzel}\ \emph {et~al.}(2018)\citenamefont {Witzel},
  \citenamefont {Hasenfratz},\ and\ \citenamefont {Rebbi}}]{Witzel:2018gxm}%
  \BibitemOpen
  \bibfield  {author} {\bibinfo {author} {\bibfnamefont {O.}~\bibnamefont
  {Witzel}}, \bibinfo {author} {\bibfnamefont {A.}~\bibnamefont {Hasenfratz}},
  \ and\ \bibinfo {author} {\bibfnamefont {C.}~\bibnamefont {Rebbi}},\
  }\href@noop {} {\  (\bibinfo {year} {2018})},\ \Eprint
  {http://arxiv.org/abs/1810.01850} {arXiv:1810.01850 [hep-ph]} \BibitemShut
  {NoStop}%
\bibitem [{\citenamefont {Witzel}\ and\ \citenamefont
  {Hasenfratz}(2019)}]{Witzel:2019oej}%
  \BibitemOpen
  \bibfield  {author} {\bibinfo {author} {\bibfnamefont {O.}~\bibnamefont
  {Witzel}}\ and\ \bibinfo {author} {\bibfnamefont {A.}~\bibnamefont
  {Hasenfratz}},\ }\href@noop {} {\  (\bibinfo {year} {2019})},\ \Eprint
  {http://arxiv.org/abs/1912.12255} {arXiv:1912.12255 [hep-lat]} \BibitemShut
  {NoStop}%
\bibitem [{\citenamefont {Detmold}\ \emph {et~al.}(2019)\citenamefont
  {Detmold}, \citenamefont {Edwards}, \citenamefont {Dudek}, \citenamefont
  {Engelhardt}, \citenamefont {Lin}, \citenamefont {Meinel}, \citenamefont
  {Orginos},\ and\ \citenamefont {Shanahan}}]{Detmold:2019ghl}%
  \BibitemOpen
  \bibfield  {author} {\bibinfo {author} {\bibfnamefont {W.}~\bibnamefont
  {Detmold}}, \bibinfo {author} {\bibfnamefont {R.~G.}\ \bibnamefont
  {Edwards}}, \bibinfo {author} {\bibfnamefont {J.~J.}\ \bibnamefont {Dudek}},
  \bibinfo {author} {\bibfnamefont {M.}~\bibnamefont {Engelhardt}}, \bibinfo
  {author} {\bibfnamefont {H.-W.}\ \bibnamefont {Lin}}, \bibinfo {author}
  {\bibfnamefont {S.}~\bibnamefont {Meinel}}, \bibinfo {author} {\bibfnamefont
  {K.}~\bibnamefont {Orginos}}, \ and\ \bibinfo {author} {\bibfnamefont
  {P.}~\bibnamefont {Shanahan}} (\bibinfo {collaboration} {USQCD}),\ }\href
  {\doibase 10.1140/epja/i2019-12902-4} {\bibfield  {journal} {\bibinfo
  {journal} {Eur. Phys. J.}\ }\textbf {\bibinfo {volume} {A55}},\ \bibinfo
  {pages} {193} (\bibinfo {year} {2019})},\ \Eprint
  {http://arxiv.org/abs/1904.09512} {arXiv:1904.09512 [hep-lat]} \BibitemShut
  {NoStop}%
\bibitem [{\citenamefont {Hasenfratz}\ \emph {et~al.}(2020)\citenamefont
  {Hasenfratz}, \citenamefont {Rebbi},\ and\ \citenamefont
  {Witzel}}]{Nf10stepScaling}%
  \BibitemOpen
  \bibfield  {author} {\bibinfo {author} {\bibfnamefont {A.}~\bibnamefont
  {Hasenfratz}}, \bibinfo {author} {\bibfnamefont {C.}~\bibnamefont {Rebbi}}, \
  and\ \bibinfo {author} {\bibfnamefont {O.}~\bibnamefont {Witzel}},\
  }\href@noop {} {\enquote {\bibinfo {title} {{Gradient flow step-scaling
  function for SU(3) with ten fundamental flavors}},}\ } (\bibinfo {year}
  {2020}),\ \bibinfo {note} {to appear in the same arXiv posting}\BibitemShut
  {NoStop}%
\bibitem [{\citenamefont {Morningstar}\ and\ \citenamefont
  {Peardon}(2004)}]{Morningstar:2003gk}%
  \BibitemOpen
  \bibfield  {author} {\bibinfo {author} {\bibfnamefont {C.}~\bibnamefont
  {Morningstar}}\ and\ \bibinfo {author} {\bibfnamefont {M.~J.}\ \bibnamefont
  {Peardon}},\ }\href {\doibase 10.1103/PhysRevD.69.054501} {\bibfield
  {journal} {\bibinfo  {journal} {Phys. Rev.}\ }\textbf {\bibinfo {volume}
  {D69}},\ \bibinfo {pages} {054501} (\bibinfo {year} {2004})},\ \Eprint
  {http://arxiv.org/abs/hep-lat/0311018} {arXiv:hep-lat/0311018 [hep-lat]}
  \BibitemShut {NoStop}%
\bibitem [{\citenamefont {Shamir}(1993)}]{Shamir:1993zy}%
  \BibitemOpen
  \bibfield  {author} {\bibinfo {author} {\bibfnamefont {Y.}~\bibnamefont
  {Shamir}},\ }\href {\doibase 10.1016/0550-3213(93)90162-I} {\bibfield
  {journal} {\bibinfo  {journal} {Nucl. Phys.}\ }\textbf {\bibinfo {volume}
  {B406}},\ \bibinfo {pages} {90} (\bibinfo {year} {1993})},\ \Eprint
  {http://arxiv.org/abs/hep-lat/9303005} {arXiv:hep-lat/9303005} \BibitemShut
  {NoStop}%
\bibitem [{\citenamefont {Furman}\ and\ \citenamefont
  {Shamir}(1995)}]{Furman:1994ky}%
  \BibitemOpen
  \bibfield  {author} {\bibinfo {author} {\bibfnamefont {V.}~\bibnamefont
  {Furman}}\ and\ \bibinfo {author} {\bibfnamefont {Y.}~\bibnamefont
  {Shamir}},\ }\href {\doibase 10.1016/0550-3213(95)00031-M} {\bibfield
  {journal} {\bibinfo  {journal} {Nucl. Phys.}\ }\textbf {\bibinfo {volume}
  {B439}},\ \bibinfo {pages} {54} (\bibinfo {year} {1995})},\ \Eprint
  {http://arxiv.org/abs/hep-lat/9405004} {arXiv:hep-lat/9405004} \BibitemShut
  {NoStop}%
\bibitem [{\citenamefont {Brower}\ \emph {et~al.}(2017)\citenamefont {Brower},
  \citenamefont {Neff},\ and\ \citenamefont {Orginos}}]{Brower:2012vk}%
  \BibitemOpen
  \bibfield  {author} {\bibinfo {author} {\bibfnamefont {R.~C.}\ \bibnamefont
  {Brower}}, \bibinfo {author} {\bibfnamefont {H.}~\bibnamefont {Neff}}, \ and\
  \bibinfo {author} {\bibfnamefont {K.}~\bibnamefont {Orginos}},\ }\href
  {\doibase 10.1016/j.cpc.2017.01.024} {\bibfield  {journal} {\bibinfo
  {journal} {Comput. Phys. Commun.}\ }\textbf {\bibinfo {volume} {220}},\
  \bibinfo {pages} {1} (\bibinfo {year} {2017})},\ \Eprint
  {http://arxiv.org/abs/1206.5214} {arXiv:1206.5214 [hep-lat]} \BibitemShut
  {NoStop}%
\bibitem [{\citenamefont {L{\"u}scher}\ and\ \citenamefont
  {Weisz}(1985{\natexlab{a}})}]{Luscher:1985zq}%
  \BibitemOpen
  \bibfield  {author} {\bibinfo {author} {\bibfnamefont {M.}~\bibnamefont
  {L{\"u}scher}}\ and\ \bibinfo {author} {\bibfnamefont {P.}~\bibnamefont
  {Weisz}},\ }\href {\doibase 10.1016/0370-2693(85)90966-9} {\bibfield
  {journal} {\bibinfo  {journal} {Phys. Lett.}\ }\textbf {\bibinfo {volume}
  {158B}},\ \bibinfo {pages} {250} (\bibinfo {year}
  {1985}{\natexlab{a}})}\BibitemShut {NoStop}%
\bibitem [{\citenamefont {L{\"u}scher}\ and\ \citenamefont
  {Weisz}(1985{\natexlab{b}})}]{Luscher:1984xn}%
  \BibitemOpen
  \bibfield  {author} {\bibinfo {author} {\bibfnamefont {M.}~\bibnamefont
  {L{\"u}scher}}\ and\ \bibinfo {author} {\bibfnamefont {P.}~\bibnamefont
  {Weisz}},\ }\href {\doibase 10.1007/BF01206178} {\bibfield  {journal}
  {\bibinfo  {journal} {Commun. Math. Phys.}\ }\textbf {\bibinfo {volume}
  {97}},\ \bibinfo {pages} {59} (\bibinfo {year} {1985}{\natexlab{b}})},\
  \bibinfo {note} {[Erratum: Commun. Math. Phys.98,433(1985)]}\BibitemShut
  {NoStop}%
\bibitem [{\citenamefont {Boyle}\ \emph {et~al.}(2015)\citenamefont {Boyle},
  \citenamefont {Yamaguchi}, \citenamefont {Cossu},\ and\ \citenamefont
  {Portelli}}]{Boyle:2015tjk}%
  \BibitemOpen
  \bibfield  {author} {\bibinfo {author} {\bibfnamefont {P.}~\bibnamefont
  {Boyle}}, \bibinfo {author} {\bibfnamefont {A.}~\bibnamefont {Yamaguchi}},
  \bibinfo {author} {\bibfnamefont {G.}~\bibnamefont {Cossu}}, \ and\ \bibinfo
  {author} {\bibfnamefont {A.}~\bibnamefont {Portelli}},\ }\href {\doibase
  10.22323/1.251.0023} {\bibfield  {journal} {\bibinfo  {journal} {PoS}\
  }\textbf {\bibinfo {volume} {LATTICE2015}},\ \bibinfo {pages} {023} (\bibinfo
  {year} {2015})},\ \Eprint {http://arxiv.org/abs/1512.03487} {arXiv:1512.03487
  [hep-lat]} \BibitemShut {NoStop}%
\bibitem [{\citenamefont {Duane}\ \emph {et~al.}(1987)\citenamefont {Duane},
  \citenamefont {Kennedy}, \citenamefont {Pendleton},\ and\ \citenamefont
  {Roweth}}]{Duane:1987de}%
  \BibitemOpen
  \bibfield  {author} {\bibinfo {author} {\bibfnamefont {S.}~\bibnamefont
  {Duane}}, \bibinfo {author} {\bibfnamefont {A.}~\bibnamefont {Kennedy}},
  \bibinfo {author} {\bibfnamefont {B.}~\bibnamefont {Pendleton}}, \ and\
  \bibinfo {author} {\bibfnamefont {D.}~\bibnamefont {Roweth}},\ }\href
  {\doibase 10.1016/0370-2693(87)91197-X} {\bibfield  {journal} {\bibinfo
  {journal} {Phys.Lett.}\ }\textbf {\bibinfo {volume} {B195}},\ \bibinfo
  {pages} {216} (\bibinfo {year} {1987})}\BibitemShut {NoStop}%
\bibitem [{\citenamefont {Wolff}(2004)}]{Wolff:2003sm}%
  \BibitemOpen
  \bibfield  {author} {\bibinfo {author} {\bibfnamefont {U.}~\bibnamefont
  {Wolff}} (\bibinfo {collaboration} {ALPHA}),\ }\href {\doibase
  10.1016/S0010-4655(03)00467-3, 10.1016/j.cpc.2006.12.001} {\bibfield
  {journal} {\bibinfo  {journal} {Comput.Phys.Commun.}\ }\textbf {\bibinfo
  {volume} {156}},\ \bibinfo {pages} {143} (\bibinfo {year} {2004})},\ \Eprint
  {http://arxiv.org/abs/hep-lat/0306017} {arXiv:hep-lat/0306017 [hep-lat]}
  \BibitemShut {NoStop}%
\bibitem [{\citenamefont {Hasenfratz}\ and\ \citenamefont
  {Knechtli}(2001)}]{Hasenfratz:2001hp}%
  \BibitemOpen
  \bibfield  {author} {\bibinfo {author} {\bibfnamefont {A.}~\bibnamefont
  {Hasenfratz}}\ and\ \bibinfo {author} {\bibfnamefont {F.}~\bibnamefont
  {Knechtli}},\ }\href {\doibase 10.1103/PhysRevD.64.034504} {\bibfield
  {journal} {\bibinfo  {journal} {Phys. Rev.}\ }\textbf {\bibinfo {volume}
  {D64}},\ \bibinfo {pages} {034504} (\bibinfo {year} {2001})},\ \Eprint
  {http://arxiv.org/abs/hep-lat/0103029} {arXiv:hep-lat/0103029} \BibitemShut
  {NoStop}%
\bibitem [{\citenamefont {Hasenfratz}\ \emph {et~al.}(2007)\citenamefont
  {Hasenfratz}, \citenamefont {Hoffmann},\ and\ \citenamefont
  {Schaefer}}]{Hasenfratz:2007rf}%
  \BibitemOpen
  \bibfield  {author} {\bibinfo {author} {\bibfnamefont {A.}~\bibnamefont
  {Hasenfratz}}, \bibinfo {author} {\bibfnamefont {R.}~\bibnamefont
  {Hoffmann}}, \ and\ \bibinfo {author} {\bibfnamefont {S.}~\bibnamefont
  {Schaefer}},\ }\href {\doibase 10.1088/1126-6708/2007/05/029} {\bibfield
  {journal} {\bibinfo  {journal} {JHEP}\ }\textbf {\bibinfo {volume} {0705}},\
  \bibinfo {pages} {029} (\bibinfo {year} {2007})},\ \Eprint
  {http://arxiv.org/abs/hep-lat/0702028} {arXiv:hep-lat/0702028 [hep-lat]}
  \BibitemShut {NoStop}%
\bibitem [{\citenamefont {Appelquist}\ \emph {et~al.}(2016)\citenamefont
  {Appelquist}, \citenamefont {Brower}, \citenamefont {Fleming}, \citenamefont
  {Hasenfratz}, \citenamefont {Jin}, \citenamefont {Kiskis}, \citenamefont
  {Neil}, \citenamefont {Osborn}, \citenamefont {Rebbi}, \citenamefont
  {Rinaldi}, \citenamefont {Schaich}, \citenamefont {Vranas}, \citenamefont
  {Weinberg},\ and\ \citenamefont {Witzel}}]{Appelquist:2016viq}%
  \BibitemOpen
  \bibfield  {author} {\bibinfo {author} {\bibfnamefont {T.}~\bibnamefont
  {Appelquist}}, \bibinfo {author} {\bibfnamefont {R.}~\bibnamefont {Brower}},
  \bibinfo {author} {\bibfnamefont {G.}~\bibnamefont {Fleming}}, \bibinfo
  {author} {\bibfnamefont {A.}~\bibnamefont {Hasenfratz}}, \bibinfo {author}
  {\bibfnamefont {X.-Y.}\ \bibnamefont {Jin}}, \bibinfo {author} {\bibfnamefont
  {J.}~\bibnamefont {Kiskis}}, \bibinfo {author} {\bibfnamefont
  {E.}~\bibnamefont {Neil}}, \bibinfo {author} {\bibfnamefont {J.}~\bibnamefont
  {Osborn}}, \bibinfo {author} {\bibfnamefont {C.}~\bibnamefont {Rebbi}},
  \bibinfo {author} {\bibfnamefont {E.}~\bibnamefont {Rinaldi}}, \bibinfo
  {author} {\bibfnamefont {D.}~\bibnamefont {Schaich}}, \bibinfo {author}
  {\bibfnamefont {P.}~\bibnamefont {Vranas}}, \bibinfo {author} {\bibfnamefont
  {E.}~\bibnamefont {Weinberg}}, \ and\ \bibinfo {author} {\bibfnamefont
  {O.}~\bibnamefont {Witzel}} (\bibinfo {collaboration} {Lattice Strong
  Dynamics}),\ }\href {\doibase 10.1103/PhysRevD.93.114514} {\bibfield
  {journal} {\bibinfo  {journal} {Phys. Rev.}\ }\textbf {\bibinfo {volume}
  {D93}},\ \bibinfo {pages} {114514} (\bibinfo {year} {2016})},\ \Eprint
  {http://arxiv.org/abs/1601.04027} {arXiv:1601.04027 [hep-lat]} \BibitemShut
  {NoStop}%
\bibitem [{\citenamefont {Hasenfratz}\ \emph
  {et~al.}(2019{\natexlab{c}})\citenamefont {Hasenfratz}, \citenamefont
  {Rebbi},\ and\ \citenamefont {Witzel}}]{Hasenfratz:2018wpq}%
  \BibitemOpen
  \bibfield  {author} {\bibinfo {author} {\bibfnamefont {A.}~\bibnamefont
  {Hasenfratz}}, \bibinfo {author} {\bibfnamefont {C.}~\bibnamefont {Rebbi}}, \
  and\ \bibinfo {author} {\bibfnamefont {O.}~\bibnamefont {Witzel}},\ }\href
  {\doibase 10.22323/1.334.0306} {\bibfield  {journal} {\bibinfo  {journal}
  {PoS}\ }\textbf {\bibinfo {volume} {LATTICE2018}},\ \bibinfo {pages} {306}
  (\bibinfo {year} {2019}{\natexlab{c}})},\ \Eprint
  {http://arxiv.org/abs/1810.05176} {arXiv:1810.05176 [hep-lat]} \BibitemShut
  {NoStop}%
\bibitem [{\citenamefont {Hasenfratz}\ and\ \citenamefont
  {Witzel}(2020)}]{Hasenfratz:2019hpg}%
  \BibitemOpen
  \bibfield  {author} {\bibinfo {author} {\bibfnamefont {A.}~\bibnamefont
  {Hasenfratz}}\ and\ \bibinfo {author} {\bibfnamefont {O.}~\bibnamefont
  {Witzel}},\ }\href {\doibase 10.1103/PhysRevD.101.034514} {\bibfield
  {journal} {\bibinfo  {journal} {Phys. Rev.}\ }\textbf {\bibinfo {volume}
  {D101}},\ \bibinfo {pages} {034514} (\bibinfo {year} {2020})},\ \Eprint
  {http://arxiv.org/abs/1910.06408} {arXiv:1910.06408 [hep-lat]} \BibitemShut
  {NoStop}%
\bibitem [{\citenamefont {Hasenfratz}\ and\ \citenamefont
  {Witzel}(2019)}]{Hasenfratz:2019puu}%
  \BibitemOpen
  \bibfield  {author} {\bibinfo {author} {\bibfnamefont {A.}~\bibnamefont
  {Hasenfratz}}\ and\ \bibinfo {author} {\bibfnamefont {O.}~\bibnamefont
  {Witzel}},\ }\href@noop {} {\  (\bibinfo {year} {2019})},\ \Eprint
  {http://arxiv.org/abs/1911.11531} {arXiv:1911.11531 [hep-lat]} \BibitemShut
  {NoStop}%
\bibitem [{\citenamefont {Fodor}\ \emph
  {et~al.}(2018{\natexlab{b}})\citenamefont {Fodor}, \citenamefont {Holland},
  \citenamefont {Kuti}, \citenamefont {Nogradi},\ and\ \citenamefont
  {Wong}}]{Fodor:2017die}%
  \BibitemOpen
  \bibfield  {author} {\bibinfo {author} {\bibfnamefont {Z.}~\bibnamefont
  {Fodor}}, \bibinfo {author} {\bibfnamefont {K.}~\bibnamefont {Holland}},
  \bibinfo {author} {\bibfnamefont {J.}~\bibnamefont {Kuti}}, \bibinfo {author}
  {\bibfnamefont {D.}~\bibnamefont {Nogradi}}, \ and\ \bibinfo {author}
  {\bibfnamefont {C.~H.}\ \bibnamefont {Wong}},\ }\href {\doibase
  10.1051/epjconf/201817508027} {\bibfield  {journal} {\bibinfo  {journal} {EPJ
  Web Conf.}\ }\textbf {\bibinfo {volume} {175}},\ \bibinfo {pages} {08027}
  (\bibinfo {year} {2018}{\natexlab{b}})},\ \Eprint
  {http://arxiv.org/abs/1711.04833} {arXiv:1711.04833 [hep-lat]} \BibitemShut
  {NoStop}%
\bibitem [{\citenamefont {Anderson}\ \emph {et~al.}(2017)\citenamefont
  {Anderson}, \citenamefont {Burns}, \citenamefont {Milroy}, \citenamefont
  {Ruprecht}, \citenamefont {Hauser},\ and\ \citenamefont {Siegel}}]{UCsummit}%
  \BibitemOpen
  \bibfield  {author} {\bibinfo {author} {\bibfnamefont {J.}~\bibnamefont
  {Anderson}}, \bibinfo {author} {\bibfnamefont {P.~J.}\ \bibnamefont {Burns}},
  \bibinfo {author} {\bibfnamefont {D.}~\bibnamefont {Milroy}}, \bibinfo
  {author} {\bibfnamefont {P.}~\bibnamefont {Ruprecht}}, \bibinfo {author}
  {\bibfnamefont {T.}~\bibnamefont {Hauser}}, \ and\ \bibinfo {author}
  {\bibfnamefont {H.~J.}\ \bibnamefont {Siegel}},\ }\href {\doibase
  10.1145/3093338.3093379} {\bibfield  {journal} {\bibinfo  {journal}
  {Proceedings of PEARC17}\ } (\bibinfo {year} {2017}),\
  10.1145/3093338.3093379}\BibitemShut {NoStop}%
\bibitem [{\citenamefont {Towns}\ \emph {et~al.}(2014)\citenamefont {Towns},
  \citenamefont {Cockerill}, \citenamefont {Dahan}, \citenamefont {Foster},
  \citenamefont {Gaither}, \citenamefont {Grimshaw}, \citenamefont {Hazlewood},
  \citenamefont {Lathrop}, \citenamefont {Lifka}, \citenamefont {Peterson},
  \citenamefont {Roskies}, \citenamefont {Scott},\ and\ \citenamefont
  {Wilkins-Diehr}}]{xsede}%
  \BibitemOpen
  \bibfield  {author} {\bibinfo {author} {\bibfnamefont {J.}~\bibnamefont
  {Towns}}, \bibinfo {author} {\bibfnamefont {T.}~\bibnamefont {Cockerill}},
  \bibinfo {author} {\bibfnamefont {M.}~\bibnamefont {Dahan}}, \bibinfo
  {author} {\bibfnamefont {I.}~\bibnamefont {Foster}}, \bibinfo {author}
  {\bibfnamefont {K.}~\bibnamefont {Gaither}}, \bibinfo {author} {\bibfnamefont
  {A.}~\bibnamefont {Grimshaw}}, \bibinfo {author} {\bibfnamefont
  {V.}~\bibnamefont {Hazlewood}}, \bibinfo {author} {\bibfnamefont
  {S.}~\bibnamefont {Lathrop}}, \bibinfo {author} {\bibfnamefont
  {D.}~\bibnamefont {Lifka}}, \bibinfo {author} {\bibfnamefont {G.~D.}\
  \bibnamefont {Peterson}}, \bibinfo {author} {\bibfnamefont {R.}~\bibnamefont
  {Roskies}}, \bibinfo {author} {\bibfnamefont {J.~R.}\ \bibnamefont {Scott}},
  \ and\ \bibinfo {author} {\bibfnamefont {N.}~\bibnamefont {Wilkins-Diehr}},\
  }\href {\doibase 10.1109/MCSE.2014.80} {\bibfield  {journal} {\bibinfo
  {journal} {Computing in Science \& Engineering}\ }\textbf {\bibinfo {volume}
  {16}},\ \bibinfo {pages} {62} (\bibinfo {year} {2014})}\BibitemShut {NoStop}%
\end{thebibliography}%
\bibliographystyle{apsrev4-1} 
\end{document}